\documentclass[twocolumn]{aastex7}

\usepackage{sidecap}
\usepackage{gensymb}
\usepackage{amsmath}

\defcitealias{Mantz2512.05405}{M25}
\newcommand*{\Mtwentyfive}{\citetalias{Mantz2512.05405}}

\makeatletter
\patchcmd{\ltx@foottext}{%
  .5\textwidth\advance\hsize-18pt}{%
  \linewidth\advance\hsize-1.8em%
}{}{}
\makeatother

\received{November 11, 2025}
\revised{January 30, 2026}
\accepted{February 13, 2026}

\begin{document}

\title{Multiwavelength Characterization of a Dynamically Relaxed Cool Core Galaxy Cluster at $z=1.5$}

\author[0000-0001-7179-6198]{Anthony M. Flores}
\affiliation{Department of Physics and Astronomy, Rutgers University, 136 Frelinghuysen Rd, Piscataway, NJ 08854, USA}
\affiliation{Department of Physics, Stanford University, 382 Via Pueblo Mall, Stanford, CA 94305, USA}
\affiliation{Kavli Institute for Particle Astrophysics and Cosmology, 452 Lomita Mall, Stanford, CA 94305, USA}
\affiliation{SLAC National Accelerator Laboratory, 2575 Sand Hill Road, Menlo Park, CA 94025, USA}
\email{aflores7@stanford.edu}
\correspondingauthor{anthony.flores7@rutgers.edu}

\author[0000-0002-8031-1217]{Adam B. Mantz}
\affiliation{Kavli Institute for Particle Astrophysics and Cosmology, 452 Lomita Mall, Stanford, CA 94305, USA}
\email{amantz@stanford.edu}


\author[0000-0003-0667-5941]{Steven W. Allen}
\affiliation{Department of Physics, Stanford University, 382 Via Pueblo Mall, Stanford, CA 94305, USA}
\affiliation{Kavli Institute for Particle Astrophysics and Cosmology, 452 Lomita Mall, Stanford, CA 94305, USA}
\affiliation{SLAC National Accelerator Laboratory, 2575 Sand Hill Road, Menlo Park, CA 94025, USA}
\email{swa@stanford.edu}

\author[0000-0003-2985-9962]{R. Glenn Morris}
\affiliation{Kavli Institute for Particle Astrophysics and Cosmology, 452 Lomita Mall, Stanford, CA 94305, USA}
\affiliation{SLAC National Accelerator Laboratory, 2575 Sand Hill Road, Menlo Park, CA 94025, USA}
\email{rgm@stanford.edu}

\author[0009-0001-9176-8861]{Abigail Y. Pan}
\affiliation{Department of Physics, Stanford University, 382 Via Pueblo Mall, Stanford, CA 94305, USA}
\affiliation{Kavli Institute for Particle Astrophysics and Cosmology, 452 Lomita Mall, Stanford, CA 94305, USA}
\email{apan5@stanford.edu}

\author[0000-0003-3521-3631]{Taweewat Somboonpanyakul}
\affiliation{Kavli Institute for Particle Astrophysics and Cosmology, 452 Lomita Mall, Stanford, CA 94305, USA}
\affiliation{Department of Physics, Faculty of Science, Chulalongkorn University, 254 Phyathai Road, Patumwan, Bangkok 10330, Thailand}
\email{taweewat.s@chula.ac.th}

\author[0000-0002-2776-978X]{Haley R. Stueber}
\affiliation{Department of Physics, Stanford University, 382 Via Pueblo Mall, Stanford, CA 94305, USA}
\affiliation{Kavli Institute for Particle Astrophysics and Cosmology, 452 Lomita Mall, Stanford, CA 94305, USA}
\email{hstueber@stanford.edu}

\author[0000-0001-5226-8349]{Michael McDonald}
\affiliation{MIT Kavli Institute for Astrophysics and Space Research, Massachusetts Institute of Technology, Cambridge, MA 02139, USA}
\email{mcdonald@space.mit.edu}



\begin{abstract}

We present imaging and spectroscopic analyses of Chandra and XMM-Newton observations of ACT-CL~J0123.5$-$0428, one of the most massive, highest redshift galaxy clusters detected within the survey fields of the Atacama Cosmology Telescope. The Chandra data are sufficient to characterize the morphology of this cluster and constrain the geometrically deprojected temperature in 2 spatial bins out to $r_{2500}$, revealing a dynamically relaxed system whose temperature drops to $kT = 1.8\pm0.6$ keV in the inner $\sim40$kpc. Within this same inner radius, the surface brightness and density of the ICM is sharply peaked, and the cooling time falls to $t_\mathrm{cool}=280^{+150}_{-120}$ Myr. A novel forward-modeling analysis of the XMM data extends imaging and spectroscopic measurements of this system out to $r_{500}$, constraining the redshift to $z=1.50\pm0.03$, with a mean temperature of $kT = 7.3\pm1.1$ keV and an emission-weighted mean metallicity of $Z/Z_\odot = 0.43^{+0.46}_{-0.25}$. We also utilize the limited optical/IR photometric coverage of the cluster to characterize the properties of the brightest cluster galaxy (BCG), which is coincident with the X-ray peak. Despite the high redshift and strong cool core, the BCG exhibits no signs of recent or ongoing star formation, 
suggesting AGN feedback has been acting persistently to stem star formation since $z\sim 2.5$.
These measurements identify ACT-CL~J0123.5$-$0428 as the highest redshift, dynamically relaxed, cool core galaxy cluster discovered to date, making it a premier target for future astrophysical and cosmological studies.

\end{abstract}



\section{Introduction} \label{sec:intro}

Galaxy clusters are the pinnacle of cosmic structure formation. These gigantic systems, which grow continuously via accretion from their surrounding environments and the sporadic merging of larger substructures, have proven to be exceptionally useful for both astrophysical and cosmological studies. The matter content of galaxy clusters is approximately 5/6 dark matter, with the remaining $\sim 1/6$ being baryonic, primarily in the hot X-ray emitting intracluster medium (ICM; see \citealt{Voit0410173}; \citealt{Boehringer2010A26ARv..18..127B}; \citealt*{Allen1103.4829}; \citealt{Kravtsov1205.5556}; \citealt{Walker1810.00890}; \citealt{Allen2019book} for reviews) 

Among the broader population of galaxy clusters, the most dynamically relaxed systems -- those that have not undergone recent subcluster merger events -- are of particular interest. For these systems, within a few sound crossing times, the ICM will approach hydrostatic equilibrium, adopting symmetric (approximately ellipsoidal) configurations. This geometric simplicity makes relaxed clusters particularly powerful as cosmological probes, allowing straightforward measurements of the three-dimensional gas and total matter properties \citep*{Allen1103.4829}. Measurements of the evolution of gas mass fraction and thermodynamic properties of such clusters have been used to place powerful constraints on cosmology (\citealt{Allen0205007,Allen0405340,Allen0706.0033,Mantz1402.6212,Mantz2111.09343,Schmidt0405374,Wan2101.09389}; see also \citealt{Ettori0904.2740,Bonamente0512349,Kozmanyan1809.09560} for studies not exclusively using relaxed systems) and the properties of dark matter \citep{Amodeo1604.02163,Mantz1607.04686,Darragh-Ford2302.10931}. 

As clusters transition to this dynamically relaxed state, their atmospheres stratify, with the coolest, densest, lowest entropy and typically highest metallicity gas sinking towards the cluster core. This leads to a characteristic sharp, central peak in the X-ray surface brightness, a drop in central temperature, and, commonly, metallicity gradients. The cooling times at the centers of these so-called ``cool core" clusters are typically significantly less than a Hubble time (often of the order a few $10^8$ yr or less, with the measurements generally limited by instrument spatial resolution; e.g.\ \citealt{White9707269,Peres9805122,Allen0002506}) and, in the absence of a balancing heat source, these regions should rapidly undergo runaway cooling (see \citealt{Fabian1994ARA&A..32..277F} for a review). However, while the central galaxies of some cool core clusters indeed host large reservoirs of cold gas and vigorous, ongoing star formation \citep{Crawford9903057,Edge0106225,Cavagnolo0806.0382,Voit0806.0384,O'Dea0803.1772,Donahue1004.0529,Edge1005.1207,McDonald1508.05941,Olivares1902.09164} generally, the amounts of cold gas and star formation observed lie far below the runaway cooling expectations, with some of the most dynamically relaxed cool core systems (of which Abell 2029 is a notable low redshift example) showing no measurable cold gas buildup or associated recent star formation \citep{Martz2003.11104}. In these systems, active galactic nucleus (AGN) feedback, which inflates cavities and drives shocks, sound waves and turbulence into the surrounding intracluster medium (ICM) appears able to effectively stem cooling and prevent star formation. Remarkably, these feedback processes also appear to limit the accretion rates onto the central black holes to allow just enough heat to be deposited into the surrounding ICM to keep the central regions of cool core clusters in an approximately steady state, for periods of up to billions of years \citep{McNamara0709.2152,McNamara1204.0006,Fabian1204.4114,Zhuravleva1410.6485,Li1503.02660,Yang1605.01725,Li1611.05455,Ehlert2204.01765}. A key question at this point is when this particular form of AGN feedback, often referred to as mechanical or radio mode feedback, first began to be important in shaping the properties of cluster cool cores.

\begin{figure}
  \includegraphics[width=\columnwidth,trim={10 15 20 35},clip]{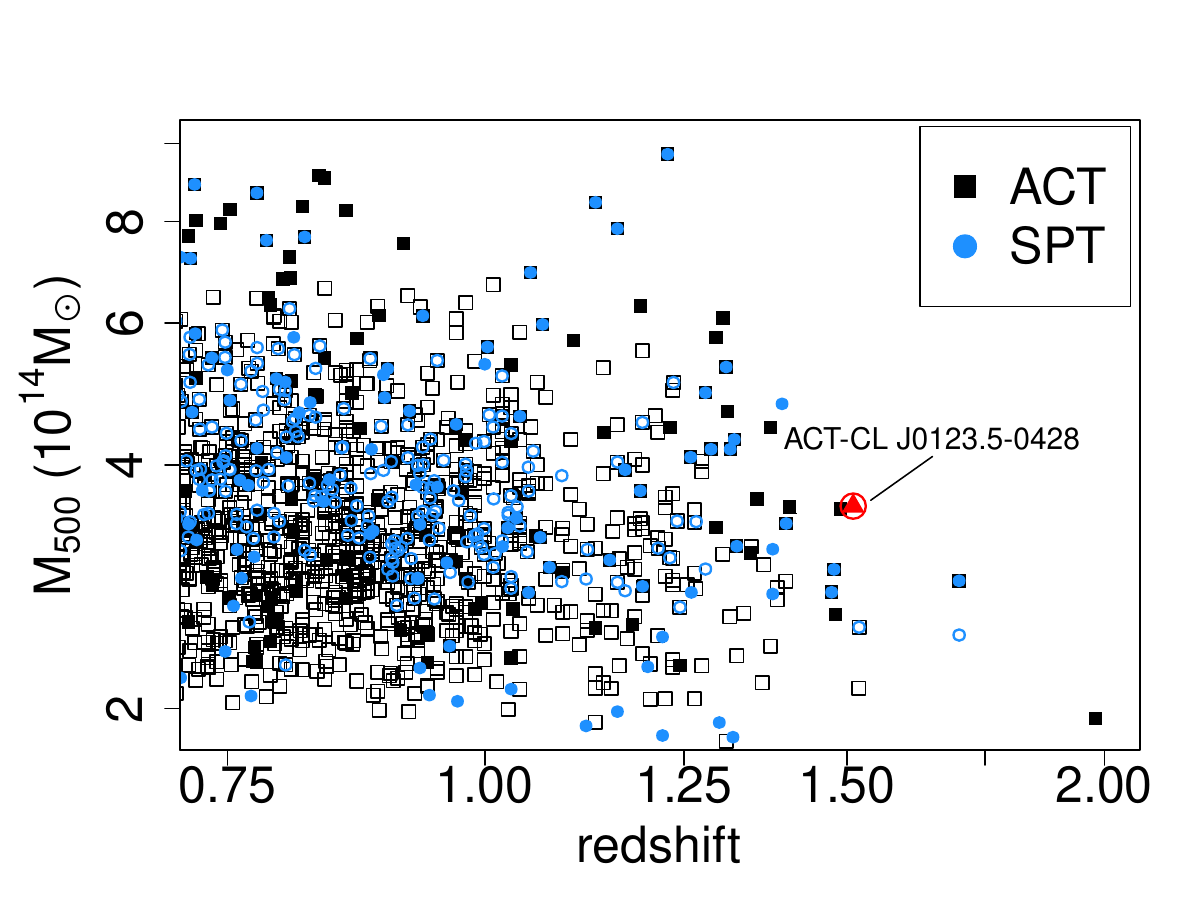}
  \caption{\small \protect\rule{0ex}{0ex}
    SZ cluster detections from the Atacama Cosmology Telescope (ACT; \citealt{Hilton2009.11043}) and the South Pole Telescope (SPT; \citealt{Bleem1409.0850,Bleem1910.04121,Huang1907.09621}), limited to $z>0.7$ and S/N~$>5$. Filled markers represent systems with targeted X-ray observations as of Chandra Cycle 27/XMM Cycle 24. 
  }
  \label{fig:SZ_sample}
\end{figure}

A key limitation affecting such studies is the dearth of confirmed dynamically relaxed, cool core clusters at high redshift. Specifically, using the quantitative X-ray morphological metrics of \citealt{Mantz1502.06020}, only 3 such systems at $z\gtrsim1$ have been previously confirmed using deep X-ray observations \citep{Mantz2111.09343}, with the most distant system, SPT-CL~J2215$-$3537, at $z=1.16$ \citep{Mantz2111.09343,Calzadilla2303.10185,Stueber2601.14425}. 
Efforts to expand these studies to higher redshifts have been challenged both by the relatively small samples of high redshift clusters known in general, and by the faint X-ray fluxes of these targets, which make detailed X-ray characterization expensive. Fortunately, the advent of cluster catalogs based on measurements of the thermal Sunyaev-Zel'dovich (SZ; \citealt{Sunyaev72}) effect is addressing the former issue. Compared to X-ray and optical measurements, the SZ effect is insensitive to the redshift of a cluster, making the detection of the most massive systems at high redshift relatively straightforward. In particular, the South Pole Telescope (SPT) and Atacama Cosmology Telescope (ACT) have observed over 5,000 and 13,000 square degrees of the southern sky, respectively, finding thousands of clusters in total, including more than 200 at $z>1$ (Figure~\ref{fig:SZ_sample}; \citealt{Bleem1409.0850,Bleem1910.04121,Huang1907.09621,Hilton2009.11043}). Although the SZ maps from SPT and ACT do not generally provide meaningful morphological information, targeted X-ray follow-up of these systems has proven an effective way to identify and characterize high mass, high redshift dynamically relaxed systems and the properties of their central ICM and galaxies (e.g.\ \citealt{McDonald1305.2915,Hlavacek-Larrondo1410.0025,Mcdonald1809.09104,Ruppin2012.14669,Ruppin2207.13351,Calzadilla2303.10185,Calzadilla2311.00396}). 


Here, we present the multiwavelength characterization of ACT-CL~J0123.5$-$0428 (hereafter ACT-CL~J0123), identifying it has the highest redshift, dynamically relaxed, cool core galaxy cluster discovered to date. With a total mass estimate of $M_{500}=3.6^{+0.7}_{-0.6}\times10^{14}M_\odot$\footnote{$M_{500}$ is defined as the mass enclosed within a sphere of radius $r_{500}$, the radius at which the mean density of the cluster exceeds the critical density of the Universe $\rho_\mathrm{crit}(z)$ by a factor of $500$. For any given overdensity, $\Delta$, M$_\Delta$ is given by: \newline $M_\Delta = \frac{4}{3}\pi\Delta\rho_{\rm crit}(z)r_\Delta^3$.} 
based on its SZ survey detection \citep{Hilton2009.11043}, ACT-CL~J0123 is 
one of the most massive, highest redshift galaxy clusters known (Figure~\ref{fig:SZ_sample}). Initially, the cluster was inferred to have a photometric redshift of $z_\mathrm{phot}=1.43\pm0.03$ from ground-based optical/near-IR survey coverage (DECaLS; \citealt{Dey1804.08657}). Here, we use X-ray measurements to determine a spectroscopic redshift of $z=1.50\pm0.03$. Our study utilizes newly developed methods for the analysis of Chandra and XMM-Newton observations to jointly characterize both the small- and large-scale physics of the X-ray emitting intracluster medium (ICM). We also utilize the aforementioned archival optical and near-IR photometry to probe the history of star formation within the central brightest cluster galaxy. 

Unless otherwise noted, all measurements are reported as the mode and associated 68.3\% credible intervals corresponding to the highest posterior probability density.  We assume a flat $\Lambda$CDM cosmology with $H_0 = 70 $ km\,s$^{-1}$\,Mpc$^{-1}$, $\Omega_\mathrm{m} = 0.3$, and $\Omega_{\Lambda} = 0.7$. At our reference redshift of $z=1.50$, the angular sizes of 1$^{\prime\prime}$ and 1${^\prime}$ respectively correspond to 8.46 kpc and 508 kpc in physical scale. Metallicities are reported relative to Solar abundance measurements of \cite{Asplund0909.0948}.

\section{Data} \label{sec:data}

ACT-CL~J0123 was observed at X-ray wavelengths as part of a joint Chandra+XMM-Newton campaign targeting the highest mass, highest redshift clusters in the Advanced ACTPol cluster catalog. 
We obtained one Chandra observation of ACT-CL~J0123 (OBSID 26940; clean exposure 51.2ks) using the ACIS-S camera. This observation is available on the Chandra Data Archive (CDA)\footnote{\href{https://cxc.harvard.edu/cda/}{https://cxc.harvard.edu/cda/}} via \dataset[doi:10.25574/cdc.529]{https://doi.org/10.25574/cdc.529}. These data were reprocessed using version 4.16 of the Chandra software analysis package, {\sc ciao}\footnote{\href{http://cxc.harvard.edu/ciao/}{http://cxc.harvard.edu/ciao/}}, and version 4.11.0 of the Chandra Calibration Database, {\sc caldb\footnote{\href{https://cxc.harvard.edu/caldb/}{https://cxc.harvard.edu/caldb/}}}. Automated lightcurve cleaning of the event file to remove periods of enhanced background was performed following the techniques recommended in the ACIS data processing guide\footnote{\href{https://cxc.harvard.edu/ciao/guides/acis_data.html}{https://cxc.harvard.edu/ciao/guides/acis\_data.html}}.

In contrast to the relatively clean Chandra observation, our XMM observations were heavily ($\sim60\%$) flared, resulting in a clean exposure time of 15/12ks in MOS/pn, respectively. These data are avalable on the ESA XMM-Newton Science Archive (OBSID 0914590601)\footnote{\href{https://nxsa.esac.esa.int/nxsa-web/}{https://nxsa.esac.esa.int/nxsa-web/}}.
The XMM data were reduced using the standard Source Analysis Software ({\sc sas}; version 18.0.0)\footnote{\href{https://www.cosmos.esa.int/web/xmm-newton/sas}{https://www.cosmos.esa.int/web/xmm-newton/sas}} tools following the guidance in the Extended Source Analysis Software ({\sc xmm-esas}) cookbook\footnote{\href{https://heasarc.gsfc.nasa.gov/docs/xmm/esas/cookbook/}{https://heasarc.gsfc.nasa.gov/docs/xmm/esas/cookbook/}}. After an initial round of automated filtering, we inspected each detector's lightcurve manually to identify and remove additional periods of flaring missed by the automated {\sc mos-filter} and {\sc pn-filter} steps. 

\begin{figure*}
\centering
  \includegraphics[scale=0.55,trim={5 0 10 0},clip]{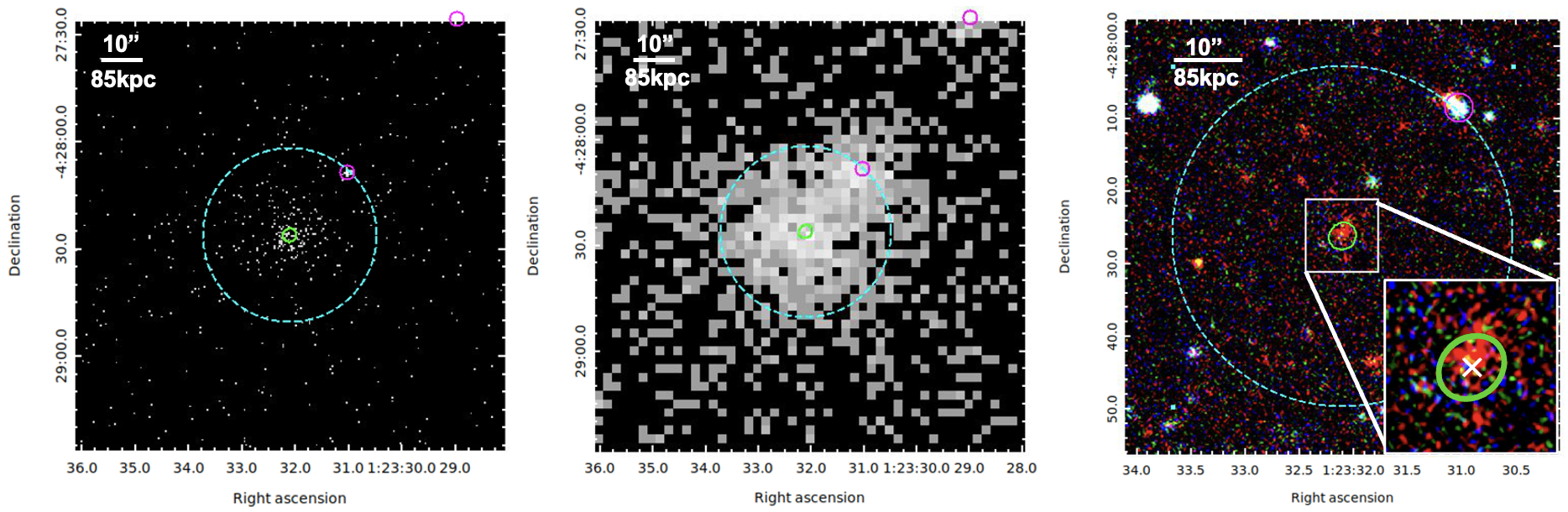}
  \caption{\small\protect\rule{0ex}{0ex}
    X-ray and optical images of ACT-CL J0123.5$-$0428. \textit{Left:} Soft band (0.6--2.0 keV) 51.2ks Chandra image. \textit{Center:} 0.4--4.0 keV stacked XMM Newton image of 15ks (MOS1/2) and 11.6ks (pn). \textit{Right}: Legacy Survey DR10 $g, r, z$ image cutout of ACT-CL~J0123. In each image, the $1\sigma$ uncertainty in the position of the cluster center (from the Chandra data) is marked with a green ellipse. Point sources identified with the CIAO tool \textsc{wavdetect} are marked with magenta circles of radius $2^{\prime\prime}$. 
    The measured vaue of r$_{2500}$ is denoted by the dashed cyan circle (See Section~\ref{sec:deprojection}). A $10^{\prime\prime}$ cutout of the center of the optical image is presented to highlight the alignment of the BCG with the X-ray cluster center (white cross). 
  }
  \label{fig:data}
\end{figure*}

The X-ray images of ACT-CL~J0123 from Chandra and XMM are shown in Figure~\ref{fig:data}. After lightcurve filtering, a simple $\beta$-model $+$ background fit to the images from each instruments reveal the Chandra observation obtained roughly 500 net cluster counts (0.6--7.0 keV), while the XMM observation obtained $\sim 700$ (0.4--4.0 keV). Over these respective energy bands, this corresponds to a measured cluster signal of $\sim50\%$ of the total background level in Chandra compared to a measured cluster signal of $\gtrsim2\times$ the total background level in XMM. 
From the Chandra image (as it is both higher resolution and exhibits less PSF blending in the cluster core compared to XMM), we obtained the cluster center from the Symmetry, Peakiness, and Alignment (SPA; \citealt{Mantz1502.06020}) algorithm, which iteratively finds the median photon position after background subtraction. We adopt this center ($\alpha = 1^{\mathrm{h}}23^{\mathrm{m}}32\overset{\mathrm{s}}{.}098$, 
$\delta = -4^{\circ}28\overset{\prime}{.}25\overset{\prime\prime}{.}320$; J2000) as the cluster center in all subsequest analyses.

Beyond these X-ray observations, ACT-CL~J0123 does not yet have targeted multiwavelength follow-up, though it does have coverage from multiple ground- and space-based surveys. The photometric redshift was determined from the Dark Energy Camera Legacy Survey (DECaLS; \citealt{Dey1804.08657}) from which it is also possible to visually identify the Brightest Cluster Galaxy (BCG), which is coincident with the center of the X-ray gas\footnote{The optical BCG position is less than 1$^{\prime\prime}$ from the SPA center, and within the SPA 1$\sigma$ uncertainties shown in Figure~\ref{fig:data}.} (see Fig~\ref{fig:data}). At this same position, coverage from the Very Large Array Sky Survey (VLASS) does not indicate any bright radio source near the location of the BCG, with epoch 3.1 observations placing a 1$\sigma$ limit on this nondetection at 69$\mu$J in the 2--4 GHz band (5--10 GHz rest frame). Similarly, data from the  Wide-field Infrared Survey Explorer (WISE) do not identify the BCG itself as an AGN, as the WISE color $(W_1 - W_2)_\mathrm{AB} = 0.06$ is too low to indicate an AGN at these redshifts \citep{Somboonpanyakul2201.08398}. ACT-CL~J0123 lacks any optical/IR spectra to precisely determine a spectroscopic redshift. To that end, the X-ray data can be used to constrain the redshift based primarily on the iron K$\alpha$ line complex around 6.7\,keV (rest frame), albeit at CCD spectral resolution. 

\section{X-ray Methods and Modeling} \label{sec:methods}
Our spectral modeling of ACT-CL~J0123 generally followed the forward modeled approach to Chandra and XMM observations described in \cite{Mantz2512.05405}; hereafter \Mtwentyfive. This approach was designed to provide the best statistical description of faint cluster environments by identifing point sources, and explicitly forward-modeling the soft foreground, unresolved X-ray background/residual AGN contamination and detector-specific non-X-ray backgrounds. Below, we briefly summarize these methods and describe their specific application to the short Chandra+XMM exposures available for ACT-CL~J0123. For our spectral analyses, all spectra were modeled using the {\sc xspec}\footnote{\href{https://heasarc.gsfc.nasa.gov/docs/xanadu/xspec/}{https://heasarc.gsfc.nasa.gov/docs/xanadu/xspec/}} analysis package (version 12.12.1c), and all thermal models are described by {\sc apec} plasma models (\citealt{Smith0106478}; ATOMDB version 3.0.9) (i.e.\ a single temperature, metallicity and density). Photoelectric absorption from gas in our own galaxy was fixed to a hydrogen column density of $3.57\times10^{20}$ cm$^{-2}$ \citep{HI4PI1610.06175}, and we used the cross sections of \cite{Balucinska1992ApJ...400..699B}. Unless otherwise stated, we constrained model parameters using the {\sc lmc}\footnote{\href{https://github.com/abmantz/lmc}{https://github.com/abmantz/lmc}} Markov Chain Monte Carlo (MCMC) code, with the likelihood given by the C-statistic in {\sc xspec} \citep{Cash1979ApJ...228..939,Arnaud1996ASPC..101...17A}. Because this approach is fully forward modeled, we were able to use the original Cash Statistic to account for the Poisson nature of the counts without grouping spectral channels.

\subsection{Point Sources} \label{sec:pnt_src}
Contaminating point sources were identified via the {\sc CIAO} tool {\sc wavdetect}\footnote{\href{https://cxc.cfa.harvard.edu/ciao/ahelp/wavdetect.html}{https://cxc.cfa.harvard.edu/ciao/ahelp/wavdetect.html}} run on broadband (0.6--7.0 keV) native resolution (0.492$^{\prime\prime}$/pixel) Chandra images. These point sources were masked in Chandra analyses using a circular mask of radius 4 times larger than the size of the the Chandra PSF corresponding to an enclosed count fraction (ECF) of 0.393\footnote{An ECF of 0.393 provides the integrated 1$\sigma$ volume of a 2-D Gaussian}. We resized these masks for XMM data based on the flux characterization returned by {\sc wavdetect} and a model for the XMM PSF \citep{Read1108.4835} to limit the residual contaminating surface brightness from each source to $\lesssim 5\times 10^{-19}$\,ergs/s/cm$^2$/arcsec$^2$ (unabsorbed 2--10 keV band) following the approach used in \cite{Mantz2006.02009} and \cite{Flores2108.12051}. For the bright AGN $\sim20^{\prime\prime}$ northwest of the cluster center (Fig~\ref{fig:data}), this default XMM mask size would have removed nearly half of all counts from the cluster. Instead, for this source, the brightest X-ray emission was masked using a $17^{\prime\prime}$ radius circle centered on the Chandra detection, with the heightened residuals from the wings of the PSF (compared to the standard procedure) accounted for in our imaging and spectral modeling (see below). For point sources in the XMM data beyond the ACIS-S footprint, we conducted a second run of {\sc wavdetect} on the 0.4--4.0\,keV pn image assuming a constant XMM PSF of 6$^{\prime\prime}$ full width at half maximum (FWHM) across the FOV. We produced circular masks with the radius given by the semi-major axis of the elliptical regions returned by {\sc wavdetect}. We found that this method produces masks that are typically just as conservative as the masks produced for regions with joint Chandra coverage.

\subsection{X-ray Spectral Methods}\label{sec:spec}
As described in \Mtwentyfive, we build and validate an initial model for various X-ray foregrounds and backgrounds (namely the soft X-ray foreground and unresolved cosmic X-ray background) by considering regions on the Chandra ACIS and XMM-Newton EPIC detectors free from cluster emission (hereafter, the ``offcluster" regions). These regions are also fitted simultaneously with cluster spectra in later stages of the analysis, largely to help constrain the strength of the various components (see below). In both instruments, 
we excluded a circle of radius 3$^\prime$ ($>2\,r_{500}$) to mask the entirety of the ICM. In Chandra, a significant portion of S3, in addition to all of CCDs S2 and S1, were free of any cluster emission and were used for the offcluster analysis. For XMM, we defined a large circular region extending 12$^\prime$ in radius, excluding the same 3$^\prime$ radius circle centered on the cluster. The outer radius was selected to encompass as much of the field of view as possible while reducing errors from systematic uncertainties due to vignetting (i.e.\ error in the calibration of the effective area at large off axis angles; \citealt{Urban1102.2430}). 

Our spectral model for the soft X-ray foreground (primarily emission from the Milky Way halo and Local Hot Bubble) comes from fits of a combination of absorbed and unabsorbed thermal models to the ROSAT All Sky Survey (RASS) maps \citep{Snowden1995ApJ...454..643S,Snowden1997ApJ...485..125S} in an offcluster region. In Chandra, we applied this model as-is, marginalizing over a 3.5\% Gaussian prior on its normalization to describe the statistical uncertainty in the ROSAT fit. In XMM, we recognized that we have superior sensitivity and spectral resolution, so the foreground model applied to XMM spectra dropped an absorbed 0.1keV thermal model in favor of the AtomDB Charge Exchange model ({\sc acx}; \citealt{Smith1406.2037}) designed to account for charge exchange in the solar wind that can contribute significantly to soft emission. Following \Mtwentyfive, we also specifically tested for the preference of a warmer ($\sim0.75$keV) absorbed thermal component, which we found ACT-CL~J0123 statistically prefers. Once the best-fitting foreground parameters (i.e.\ temperatures and normalizations of the {\sc apec/acx} components) were determined for the XMM offcluster region, they were fixed for later stages of the analysis. When applying this model simultaneously to smaller regions that include cluster emission, the normalizations were scaled by the ratio in area to the offcluster region, and an overall constant scaling was applied to all regions as a free parameter..

In addition to the residual emission from masked, detected sources due to the wings of the XMM PSF, we also predicted the contribution from the unresolved (i.e.\ not detected by Chandra) cosmic X-ray background following the same methods outlined in \Mtwentyfive. In brief, {\sc marx} simulations of sources located at increasing off-axis angles provide sensitivity limits (at the 99\% level) for the detection of point sources as a function of source flux, distance from the Chandra aimpoint (i.e.\ Chandra PSF size), and ``background'' level (including ICM emission for point sources within the cluster or in projection). Assuming an AGN X-ray luminosity function (\citealt{Miyaji1503.00056}), we obtained the total expected contribution from unresolved AGN in each pixel by integrating this luminosity function up to the flux corresponding to our sensitivity limits. In our spectral analyses, the unresolved and residual AGN components were modeled by a single absorbed powerlaw with a photon index of $1.4$. The normalization of this powerlaw in each spatial region was fixed to the total expectation from the methods described above, with an overall scaling shared between spatial regions. For XMM, a significant portion of the offcluster region was outside the Chandra FOV. Masks applied to sources identified in these data were independent of Chandra, thus our Chandra PSF-dependent sensitivity model was invalid. In this case, the normalization of the powerlaw model was not fixed to any expectation, nor was it linked to the models of other simultaneously-modeled regions.


We generated models to describe the quiescent particle background (QPB) in both telescopes, following the work of \cite{Suzuki2108.11234} for Chandra ACIS and using the ESAS {\sc mos-back} and {\sc pn-back} tools for XMM EPIC. Both methods provide an observation-specific spatially varying model of the X-ray particle background to be forward modeled into our spectral analyses. In Chandra, these models are specific to each detector, using the high-energy count rate to scale the expected spectrum. Each Chandra QPB model was marginalized over the statistical uncertainty in the normalization, with an additional 5\% allowance overall to account for additional systematics. In XMM, the ESAS QPB model spectrum is converted to an {\sc xspec} table model with fixed shape but free normalization. Instead of allowing this normalization to vary for each spatial region fit, we fixed the relative normalizations of each region given by the ESAS model expectation and introduced a single constant scaling factor specific to each detector (MOS1, MOS2, and pn). Each of these detector-specific scalings was marginalized over an independent 10\% Gaussian prior. 

Finally, we used the same methods as \Mtwentyfive\ to produce a smooth model for the pn Out-of-Time (OOT) events. The same {\sc SAS} tools that produce science spectra also provide an estimate of the shape of the OOT spectrum by resampling the events in the readout direction. We fit this spectrum with a combination of 2 absorbed powerlaws, 2 Gaussians, and the smoothed QPB model described above. One pair of powerlaw+Gaussian components is folded through the instrumental response ({\sc arf}) while the other is not. This process is not designed to describe any physical process, but instead provide an estimate of the spectral shape of OOT events (see \Mtwentyfive\ for further details and examples). The normalizations of each component in the OOT model were fixed and scaled by a constant value of 0.063 (appropriate for Full Frame mode observations) in all spectral analyses.

In Chandra, the default energy range of our fits was 0.6--7.0 keV. For regions containing cluster emission that were dominated by the particle background (i.e. at large radius), we narrowed the energy bounds of our fits to match the energies at which the cluster signal is higher than the particles. In XMM, we fit from 0.4--12.0 keV in MOS1/2 and 0.4--15.0 keV in pn, excluding 1.2--1.9 keV for both MOS detectors to remove the Al and Si fluorescence lines, and 1.2--1.65 keV in pn to remove the Al line. Similarly, we exclude 7.0--10.0 keV to exclude the Ni, Cu, and Zn fluorescence lines. While the 10.0--15.0 keV band is devoid of any cluster signal, it was especially useful to normalize the particle background and to test for the presence of residual soft protons, for which we saw no visual evidence.

\section{Results}

\subsection{Cluster Morphology} \label{sec:morph}
\begin{figure*}
\centering
  \includegraphics[scale=0.45]{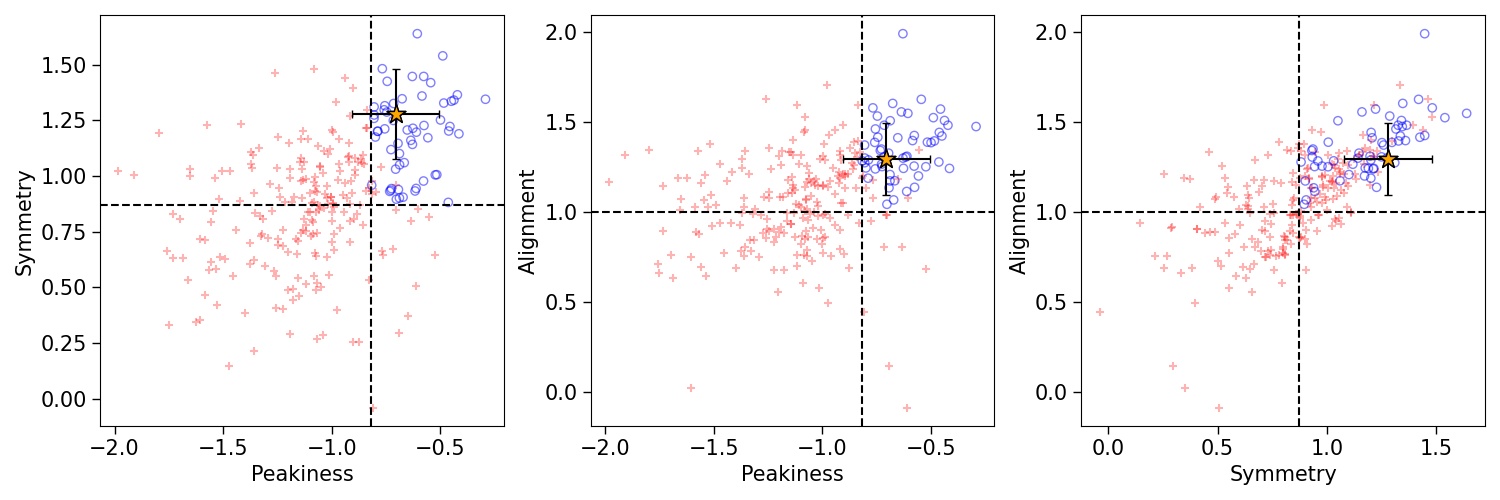}
  \caption{\small\protect\rule{0ex}{0ex}
    Symmetry, Peakiness and Alignment metrics describing the X-ray morphology of ACT-CL J0123, compared with a large sample of clusters \citep{Mantz1502.06020}. ACT-CL J0123 (orange star) satisfies the criteria for relaxation, which is defined in terms of exceeding thresholds (dashed lines) in all 3 parameters. For the larger sample, blue circles (red crosses) show clusters classified as relaxed (unrelaxed).
  }
  \label{fig:SPA_param}
\end{figure*}

At such high redshift, 
the HPD of the XMM PSF spreads emission over an area greater than 120kpc in scale, significantly complicating small-scale probes of the ICM surface brightness distribution, particularly the peakiness of the profile. 
Therefore, the physics of the core and overall morphology of ACT-CL~J0123 were entirely constrained by the Chandra data. 
To determine the morphology of ACT-CL~J0123, we computed the SPA metrics from the soft-band (0.6--2.0 keV) Chandra image and bootstrapped this computation 10,000 times to obtain an estimate of the error on each morphological parameter. The SPA metrics calculated for ACT-CL~J0123 ($s=1.3\pm0.2$, $p=-0.7\pm0.2$, and $a=1.3\pm0.2$) are plotted in Fig.~\ref{fig:SPA_param} alongside those of other dynamically relaxed and unrelaxed systems. ACT-CL~J0123 met the SPA criteria for relaxation ($>50\%$ of bootstrap measurements simultaneously satisfying $s >0.87$, $p>-0.82$, and $a>1.0$; \citealt{Mantz1502.06020}). 


\subsection{Redshift} \label{sec:single_spec}
\begin{figure*}
\centering
  \includegraphics[width=\textwidth]{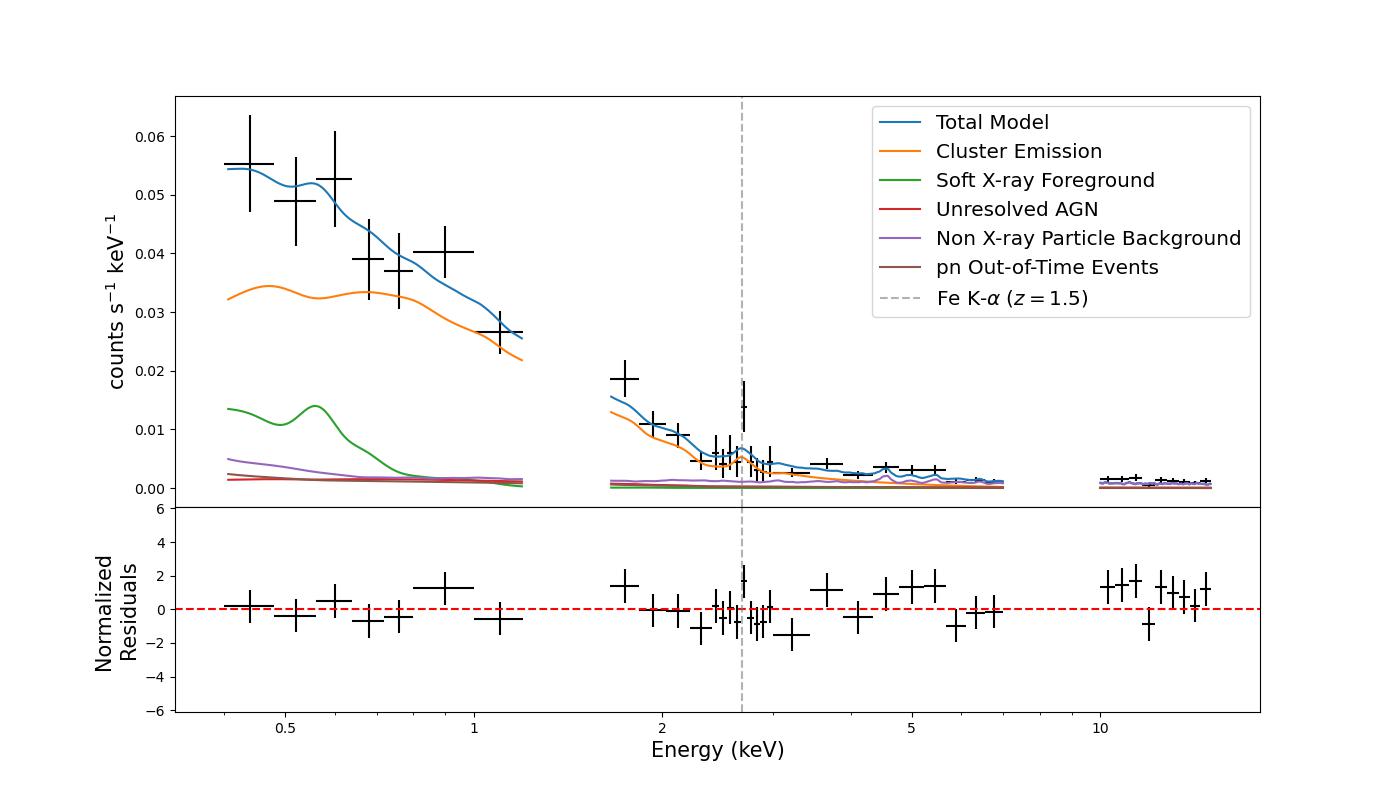}
  \caption{\small\protect\rule{0ex}{0ex}
    \textit{Top:} pn spectrum of ACT-CL~J0123, extracted from a circular region of radius 80$^{\prime\prime}$ centered on the cluster. The spectrum is binned in energy for clarity, with a finer binning around 2.7 keV (6.7 keV rest frame) to highlight the Fe K-$\alpha$ detection. The vertical dashed grey line indicates Fe K-$\alpha$ emission redshifted by $z_\mathrm{spec}=1.5$. Also plotted are the total model (blue) and individual model components, including the cluster (orange), X-ray foreground (green), cosmic X-ray background (red), quiescent particle background (QPB; purple) and pn out-of-time (OOT) events (brown). Gaps in the model curves and between data points correspond to the excluded energies in our fits due to fluorescence lines. \textit{Bottom:} Best-fit model residuals to the pn data, normalized by their uncertainty.
  }
  \label{fig:single_spec}
\end{figure*}

As ACT-CL~J0123 currently lacks optical/near-IR spectroscopic follow-up of its cluster members, we used the XMM data to constrain its redshift. We extracted a single spectrum from an 80$^{\prime\prime}$ circle surrounding the cluster center, corresponding to the extent of detected cluster emission in the XMM observation 
and fit a single absorbed thermal model ({\sc phabs$\times$apec}). 
We found the best-fit spectroscopic redshift of $z_\mathrm{spec}=1.50\pm0.03$, and fixed the value at $z=1.50$ in all later stages of analysis. The spectrum used to determine the redshift, the best fit model (including individual model components to highlight their relative strengths), and the model residuals are presented in Figure~\ref{fig:single_spec}.

\subsection{Annular Deprojection} \label{sec:deprojection}
To produce radial profiles (e.g.\ gas density or temperature), we extracted spectra in concentric annuli surrounding the cluster center. The modeling of these data accounted for both the geometric effect of the projection of emission from spherical shells onto circular annuli of smaller radius and the mixing of emission between annuli due to the instrumental PSF. 
\citep{Fabian1981ApJ...248...47F, Kriss1983ApJ...272..439K}. 
In our Chandra analysis, this took the form of the {\sc projct} model which accounts for the geometric mixing ignoring the negligible effects of the Chandra PSF. In XMM where the mixing from the PSF is significantly more pronounced, we explicitly calculated a matrix of mixing values that accounts for effects of both geometric projection and the PSF to describe how the emission from a given annulus is spread into all others. To predict the spread of cluster emission by the PSF, we used the best-fit parameters of a $\beta-$model fit to the Chandra data where PSF mixing is minimal \citep{Mantz2006.02009}. 
To obtain the gas density profile for the cluster, we converted the model normalizations\footnote{The normalization of the {\sc apec} model is defined as $\frac{10^{-14}}{4\pi D_A\left(1+z\right)^2}\int n_e n_HdV$ and is therefore a proxy for the emissivity of the gas from which we can recover the gas electron density.} to ICM density using the canonical mean molecular mass of $\mu = 0.61m_p$, our reference cosmology, and the cluster redshift $z=1.5$.
The final bin of the Chandra and XMM profiles is corrected for potential emission mixing in from beyond the outermost spectral region considered according to the procedure in \Mtwentyfive. 

\begin{figure*}
\centering
  \includegraphics[scale=0.388]{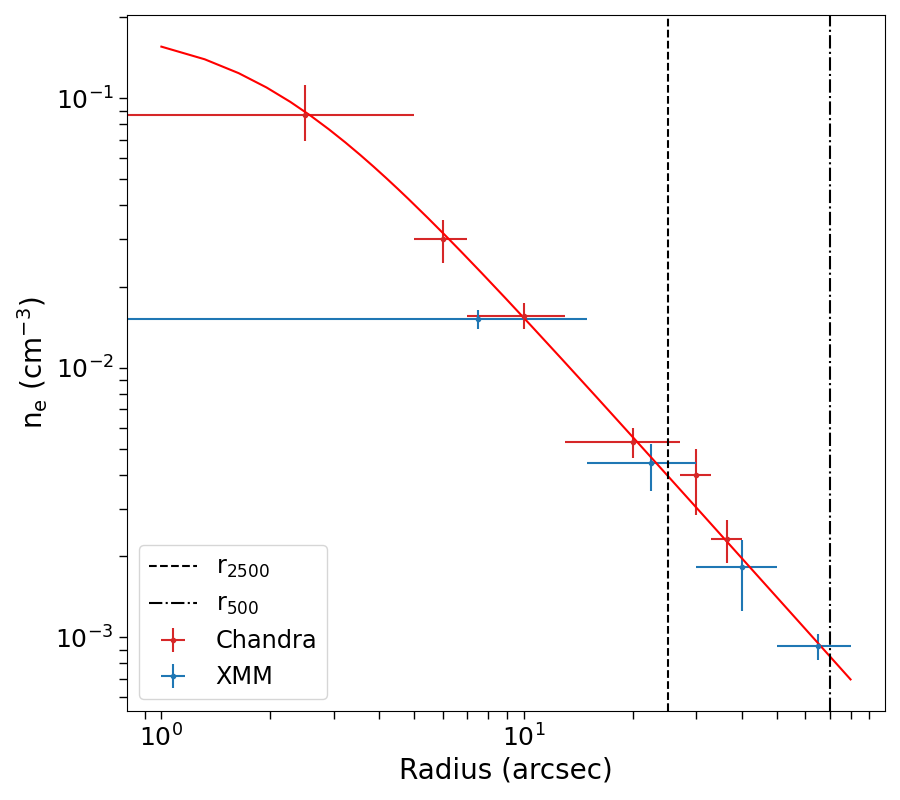}
  \includegraphics[scale=0.385]{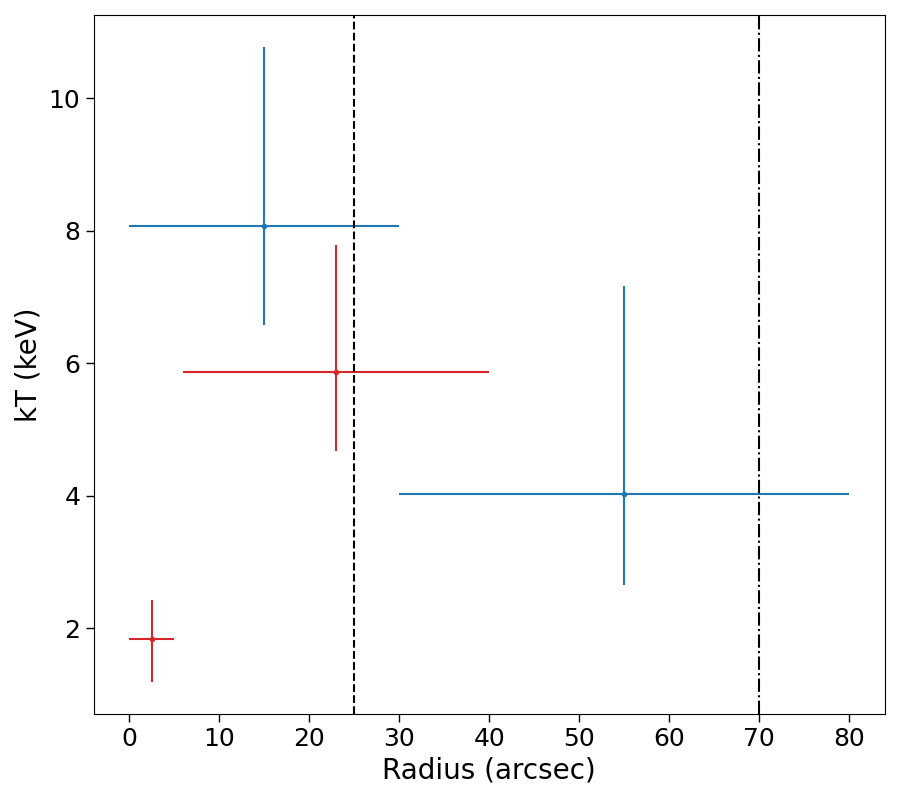}
  \caption{\small\protect\rule{0ex}{0ex}
    Density (left) and temperature (right) profiles for ACT-CL~J0123 as measured by Chandra (red) and XMM (blue). Chandra is able to resolve more of the inner density profile, demonstrating its peakiness, and identifies the cool core. A $\beta$-model fit to the Chandra density (solid red curve) shows good agreement with XMM beyond the central bin, even at large radiii where Chandra cannot probe.
  }
  \label{fig:profile_telescope_comp}
\end{figure*}

We present the profiles of electron density and temperature from Chandra and XMM in Fig~\ref{fig:profile_telescope_comp}. 
The Chandra data are able to more finely resolve the center of the cluster, as demonstrated by the sharply peaked density and the identification of the cool core. Beyond the central XMM bin, where the impact of the larger PSF was substantial, we saw very good agreement in the radial density profiles. Using these profiles, we also computed the X-ray Pseudo-Entropy ($K = k_BT_X \times n_e^{-2/3}$) and cooling time 
\begin{equation}
t_\mathrm{cool} = \frac{3}{2}\frac{(n_e+n_p)kT}{n_e n_p \Lambda(T,Z)},
\end{equation}
where $\Lambda(T,Z)$ is the cooling function of the gas. Notably, the temperature of the gas in the inner 40kpc of the cluster was constrained to be $1.8\pm0.6$ keV, with a corresponding central entropy of $9\pm4$ keV cm$^2$ and cooling time of $t_\mathrm{cool}=280^{+150}_{-120}$ Myr. 
Neither telescope could constrain more than a single, spatially-averaged, emission-weighted metallicity, with $Z_\mathrm{XMM}=0.43^{+0.46}_{-0.25}$ and $Z_\mathrm{Chandra}=0.31\pm0.25$, respectively. Here, we have applied the cross-calibration factor of $\ln \left(Z/Z^\mathrm{Cha}\right) = 0.28^{+0.10}_{-0.07}$ to the Chandra metallicity from \cite{Flores2108.12051} which provides better agreement between Chandra-measured abundances and those of Suzaku and XMM.

\subsection{Mass Estimates}\label{sec:mass}
We integrated the deprojected radial density profiles from Sec~\ref{sec:deprojection} 
and used them to simultaneously constrain $r_\Delta$ and $M_\Delta$ for $\Delta = 2500$ and $500$, accounting for the correlated measurement uncertainties in density as a function of radius. 
When calculating total mass, we assumed fiducial $f_{\rm gas}=M_\mathrm{gas}/M_\mathrm{tot}$ values of 0.113 at $r_{2500}$ and 0.125 at $r_{500}$ based on measurements of similar clusters at $z\leq1.063$ \citep{Allen0706.0033,Mantz1509.01322}. We note that the gas mass fraction at these overdenstities is not expected to evolve significantly with redshift for large halos based on cosmological scale hydrodynamical simulations (e.g.\ \citealt{Eke9708070,Nagai0609247,Battaglia1209.4082,Planelles1209.5058,Barnes1607.04569,Singh1911.05751,Rasia2505.21624,Aljamal2507.05176}). For ACT-CL~0123, we measured $r_{2500} = 206\pm14$kpc ($\sim25^{\prime\prime}$) with an enclosed mass of $M_{2500} = 6.7\pm1.3\times10^{13}M_\odot$. Only by using the XMM data could we reach radii larger than $r>r_{500}$, yielding $r_{500}=590\pm40$kpc ($\sim70^{\prime\prime}$) and $M_{500}=3.1\pm0.6\times10^{14}M_\odot$. 

\subsection{BCG Photometric Modeling}\label{sec:sfr}
\begin{figure*}
\centering
  \includegraphics[scale=0.4]{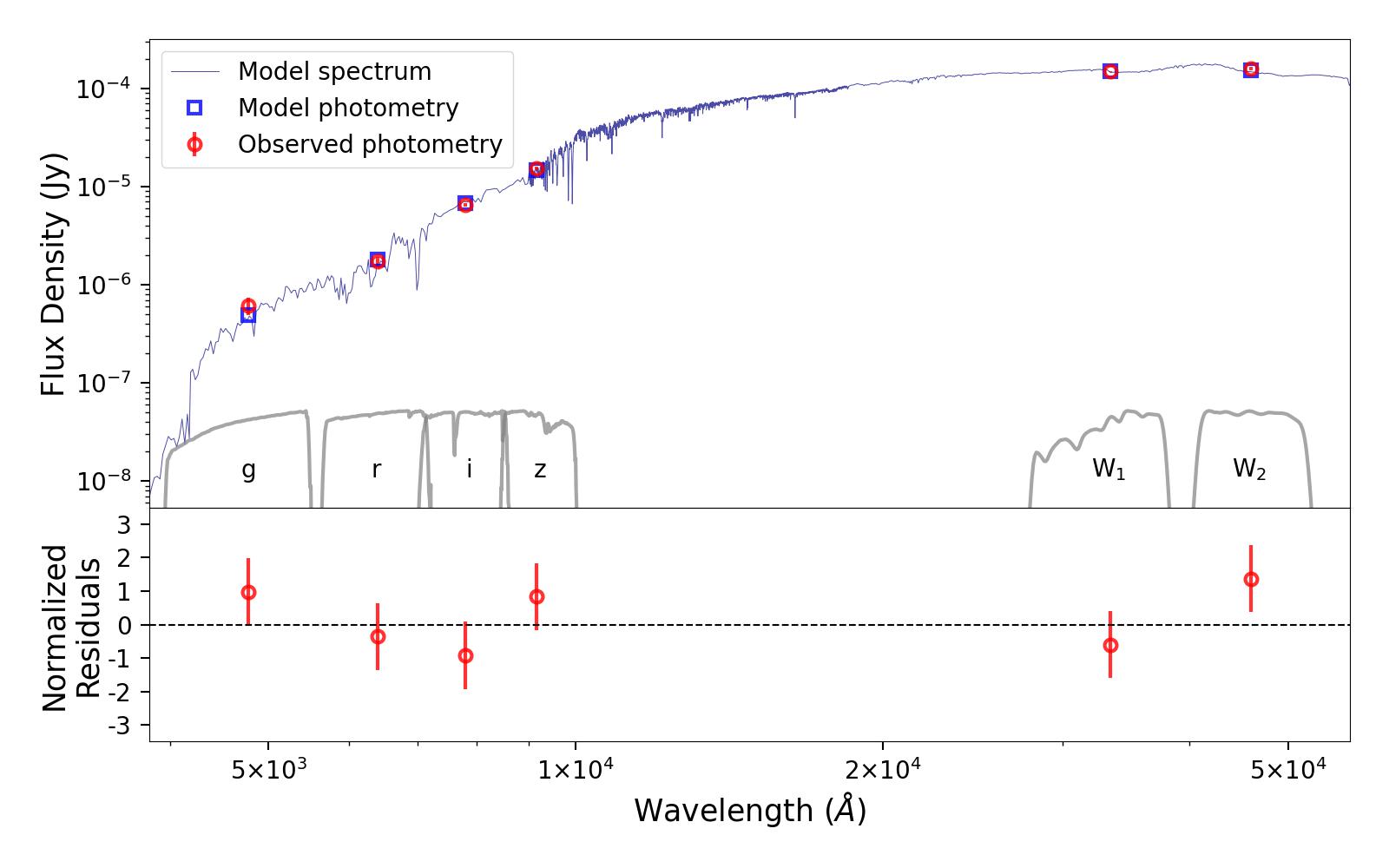}
  \caption{\small\protect\rule{0ex}{0ex}
    \textit{Top:} Best fitting simple stellar population (i.e.\ starburst) model spectrum to observed-frame optical (DECaLS $g,r,i,z$) and near-IR (WISE $W_1$ and $W_2$) photometry for the BCG of ACT-CL~J0123. The photometric data with error bars are plotted in red. The relative transmission curves for each photometric band are plotted in gray and labeled by name. Using a simple stellar population model (SSP) in {\sc prospector}, we obtain the best-fitting SED and corresponding model photometry (blue curve and squares, respectively). \textit{Bottom:} Normalized residuals for the best fit SSP model. This spectrum points toward the BCG hosting an evolved stellar population, with no evidence for active star formation for more than a Gyr before the time of observation.
  }
  \label{fig:prospector}
\end{figure*}

ACT-CL~J0123 lacks any detailed spectroscopic optical/near-IR follow-up of its galaxy population. In lieu of this, we combined optical photometry of the cluster BCG in the $g,r,i,z$ bands from DR10 of the Legacy Survey (DECaLS; \citealt{Dey1804.08657}) with $W_1, W_2$ near-IR measurements from the CatWISE (AllWISE + NEOWISE; \citealt{Eisenhardt1908.08902}) catalog constructed using observations from the Wide Field Infrared Explorer (WISE). We used the {\sc extinction} package \citep{barbary_2016_804967} to correct for local extinction/reddening given by the \cite{Cardelli1989} Milky Way model, fixed to the local value of $A_V = 0.0869$ \citep{Schlafly1012.4804}. Using the {\sc prospector} code \citep{Johnson2012.01426}, we fit these 6 corrected photometric points using a Simple Stellar Population model (SSP; \citealt{Conroy0809.4261,Conroy0911.3151}) to simultaneously constrain the total formed stellar mass, metallicity, and age of the BCG, with an additional free parameter to account for intrinsic dust extinction following the \cite*{Kriek1308.1099} attenuation curves. In these fits, we assumed the initial mass function (IMF) from \cite{Kroupa0009005}. We obtained posterior distributions for these parameters by running MCMC chains using the {\sc emcee} package \citep{Foreman-Mackey1202.3665}. The resulting spectrum is plotted in Fig~\ref{fig:prospector}. 
Our photometric analysis of the BCG yielded a stellar metallicity of $\log_{10}\left(Z/Z_\odot\right) = -0.19^{+0.14}_{-0.28}$, a total formed stellar mass of $M_*=8.6^{+2.3}_{-1.3}\times 10^{11} M_\odot$, and a galactic age of $t_\mathrm{age}=1.5^{+0.4}_{-0.9}$ Gyr. The corresponding ``surviving" galactic stellar mass (i.e. the galactic stellar mass at the time of observation) was found to be $M_*=4.9^{+1.6}_{-0.9}\times 10^{11} M_\odot$).
Dust attenuation (defined as opacity at 5500$\mathrm{\AA}$) was constrained to be $0.11^{+0.13}_{-0.10}$. In addition to this choice of model, we tested models neglecting intrinsic dust extinction (i.e.\ fixing opacity to 0) and found strong statistical evidence that dust should be included as a free parameter. We also tested whether a delayed star formation model is preferred to an instantaneous starburst, but found no statistical preference for this more complex model (whether dust was included or not) and no significant differences in the posteriors of common fitted parameters. These data suggest that there has been no significant star formation in the BCG for the past $\sim1.5$ Gyr (when the Universe was $\sim 2.7$ Gyr old), with the bulk of the stellar mass formed at a redshift of $2.4^{+0.4}_{-0.6}$.

\begin{figure*}
\centering
  \includegraphics[scale=0.8]{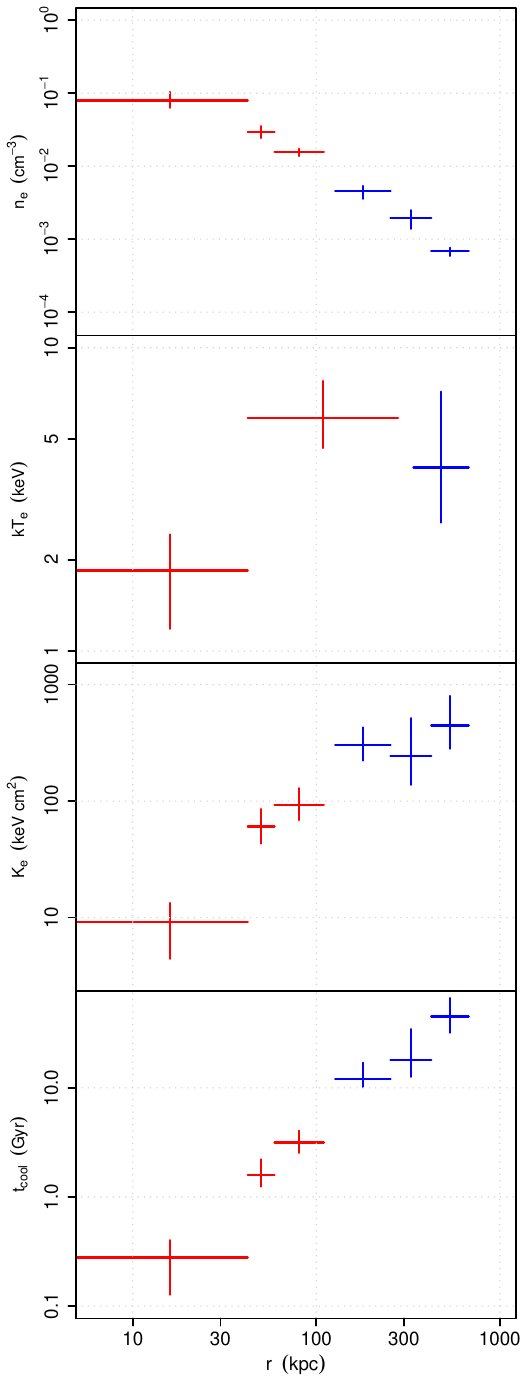}
  \includegraphics[scale=0.8]{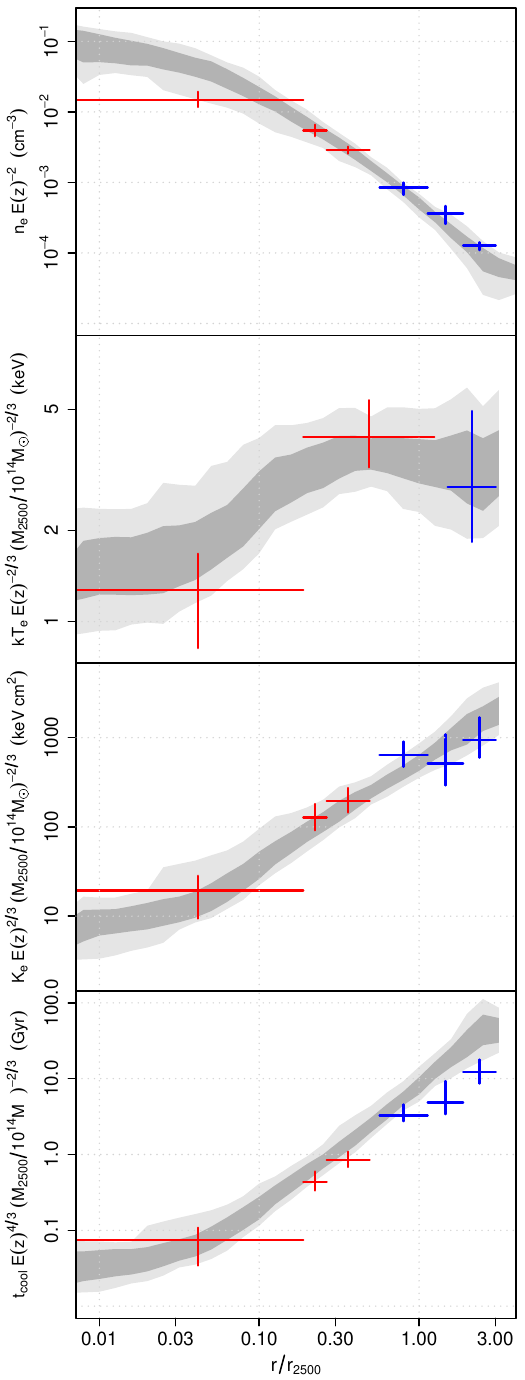}
  \caption{\small\protect\rule{0ex}{0ex}
    Thermodynamic profiles for ICM electron density, temperature, pseudoentropy, and cooling time (top $\rightarrow$ bottom) of relaxed clusters. ACT-CL~J0123 is constrained at small radii by Chandra (red points) and at large radii by XMM-Newton (blue points). Profiles of ACT-CL~J0123 alone are presented on the left, unscaled, with the radius in physical units. Profiles on the right include self-similar scaling factors, and are compared with an ensemble of 40 relaxed systems (grey band) previously analyzed in \cite{Mantz1509.01322}.
  }
  \label{fig:thermo}
\end{figure*}

\begin{figure}
\centering
  \includegraphics[width=\columnwidth]{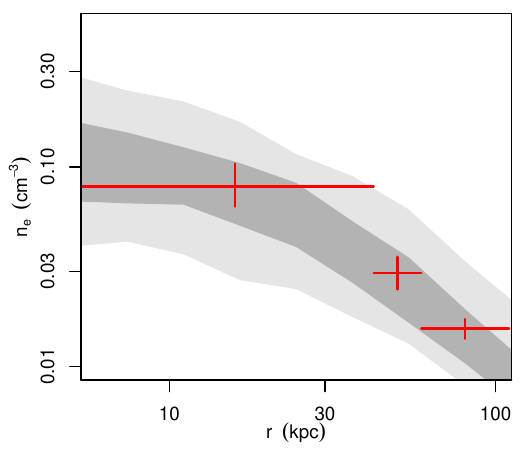}
  \caption{\small\protect\rule{0ex}{0ex}
    Central ICM density profile of ACT-CL~J0123 compared to the sample of relaxed clusters from \cite{Mantz1509.01322}, without self-similar scaling (c.f.\ Fig~\ref{fig:thermo}). 
  }
  \label{fig:dens_unscal}
\end{figure}


\section{Discussion}\label{sec:discussion}
Based on the imaging and spectroscopic results outlined here, ACT-CL~J0123 is the highest-redshift dynamically relaxed, cool core cluster discovered to date. 
The SPA morphological tests in this work were dependent on the Chandra data, as XMM-Newton could not adequately probe the cluster core with its poorer spatial resolution. In principle, XMM data could help characterize ICM dynamics of high-redshift clusters at more intermediate radii (see \citealt{Bartalucci1610.01899}), but in this case this would be complicated by the nearby point source (Fig~\ref{fig:data}) and the diminished azimuthal coverage and statistics as a result of the required masking.

Despite the ICM reaching low entropies in the core, a feature often associated with star formation in the BCGs of cool core clusters \citep{Cavagnolo0806.0382,Voit0806.0384,Calzadilla2311.00396}, we saw no evidence for ongoing or recent star formation in the BCG of ACT-CL~J0123 based on currently available optical/near-IR photometry (Sec~\ref{sec:sfr}). In contrast, the BCG of the next highest redshift relaxed cluster, SPT-CL~J2215$-$3537 ($z=1.16$; \citealt{Calzadilla2303.10185}), exhibits a starburst of $>300\ \mathrm{M}_\odot \ \mathrm{yr}^{-1}$. However, not all high-redshift relaxed systems demonstrate high levels of star formation. At slightly lower redshift ($z=1.02$), the central galaxy of relaxed cluster WARP~J1415.1$+$3612 
also exhibits a low core entropy ($9.9\pm2.0$ keV cm$^2$), with an upper limit on the star formation rate of its BCG of $<10\ \mathrm{M}_\odot \ \mathrm{yr}^{-1}$ (two orders of magnitude lower than predicted by the cooling rate; \citealt{Santos1111.3642}). At $z=1.40$, SPT-CL~J0607$-$4448 is another example of a system with a comparatively high central gas density (and relatively low central entropy) for which a tight limit on the star formation rate in its BCG of only $\sim 1 \mathrm{M}_\odot \ \mathrm{yr}^{-1}$ has ben obtained \citep{Masterson2301.00830}.%
\footnote{However, SPT-CL~J0607$-$4448 does not exhibit a cool core with a central temperature drop in its central 50\,kpc, nor is surface brightness ``peaky'' enough for the cluster to meet the SPA criteria for dynamical relaxation.}

The canonical explanation for quenched star formation in cool core galaxy clusters is AGN feedback, which provides heat to offset runaway cooling. It is possible for the central galaxy of relaxed, high redshift clusters to be remarkably active; at $z=1.06$, 3C~186 represents an extreme example, where the BCG is an exceptionally bright radio-loud quasar \citep{Siemiginowska1008.1739}. However, for ACT-CL~J0123, the current non-detection of a central X-ray point source and lack of AGN signatures in the IR (CatWISE) and existing radio data (VLASS) argue against such a possibility. We do note a slight deficit of counts to the SW of the cluster center in the native-resolution Chandra image, which may be indicative of recent jet-ICM interactions (Fig~\ref{fig:data}); however, the existing data are insufficient to further explore the possibility of cavities in the intracluster gas at present.

Our photometric analysis of the BCG suggests that there has been no significant star formation in the BCG for the past $\sim1.5$ Gyr. However, we note that the existing data are somewhat limiting. Currently, the only photometry redward of the 4000$\mathrm{\AA}$ break come from WISE, whose large PSF ($\sim 6^{\prime\prime}$ in W$_1$ and W$_2$) could result in significant blending. This issue is mitigated somewhat by the forced photometry employed for the DECaLS catalogues  \citep{Dey1804.08657}, but these fits would be significantly enhanced by additional near-IR imaging or spectroscopy at higher spatial resolution (e.g.\ from JWST). In our fits, we see significant degeneracy between dust attenuation and the age of the last starburst, with higher opacity corresponding to a more recent starburst. Free intrinsic extinction is greatly statistically preferred; however, simply including this free parameter reduces the best-fit stellar age by $\sim1$ Gyr. To place the best constraints on the stellar history of this BCG (and that of other cluster members), photometry bridging the DECaLS and WISE observations would be particularly useful.

Observations of high redshift galaxy clusters are also invaluable to explore the history of cosmic enrichment. 
As the available XMM data only provided a bulk detection of metallicity within the cluster of $Z/Z_\odot=0.43^{+0.46}_{-0.25}$, we currently are unable to comment on the radial abundance distribution in the ICM. However, high redshift studies of cluster metallicity would lead us to expect the Fe abundance to drop to $0.2\lesssim Z/Z_\odot\lesssim0.3$ by $r_{500}$ \citep{Ettori1504.02107,McDonald1603.03035,Mantz1706.01476,Liu200312426,Flores2108.12051}, and it would be interesting to determine whether ACT-CL~J0123 hosts a metallicity gradient. Central metallicity gradients present at the time of cluster formation can remain in place without merger activity (or significant AGN outbursts) to distribute metals to larger radii, as seen in a variety of observations (e.g. \citealt{Allen9802219,De-Grandi0012232}) and simulations (e.g. \citealt{Rasia1509.04247}). Moreover, the cores (on $\sim1$0 kpc scales) of highly relaxed systems can reveal near-Solar metallicities \citep{Mcdonald1809.09104} or even more extreme abundances, such as the $\sim3Z_\odot$ core seen in WARP~J1415.1$+$3612 \citep{Santos1111.3642}. Even for other high-z relaxed clusters where the central metallicity is not as extreme (e.g. SPT-CL~J0615$-$5746 and SPT-CL~J2215$-$3537), core and core-excised metallicity ratios of $\sim 2$ are still observed \citep{Jimenez-Teja2305.10860,Stueber2601.14425}.


The observed thermodynamic properties of clusters as a function of redshift also provide a powerful probe of cluster evolution. In the context of the self-similar model \citep{Kaiser1986MNRAS.222..323K}, where galaxy clusters form purely from gravitational collapse, the observable properties of clusters (density, thermodynamics, luminosity, etc), should scale simply with redshift and mass. Hydrodynamical simulations 
have also been used to study the evolution of clusters, incorporating additional astrophysical processes including cooling and feedback from stars and AGN (for a review, see \citealt{Borgani0906.4370}). More recent work (including e.g.\ MACSIS, Millenium, {\sc illustris}TNG, TNG-{\sc cluster}, and {\sc flamingo}; \citealt{Battaglia1109.3709,Scott1112.3769, Barnes1607.04569,Pop2205.11528,Aljamal2507.05176}) shows that, even in the presence of feedback processes, scaling relations for the most massive halos (with $M_{500}\gtrsim10^{14} M_\odot$) should follow self-similar predictions quite closely at intermediate to large scales.  Only in the centers of such clusters, and particularly cool core systems, are deviations from self-similar predictions routinely observed \citep{Arnaud0910.1234, McDonald1404.6250, Mantz1509.01322,Mcdonald1809.09104}.

\begin{figure}
\centering
  \includegraphics[scale=1.15,trim={10 17 20 33},clip]{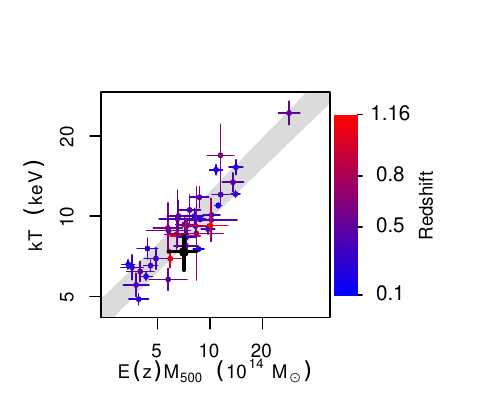}
  \includegraphics[scale=1.15,trim={10 17 20 33},clip]{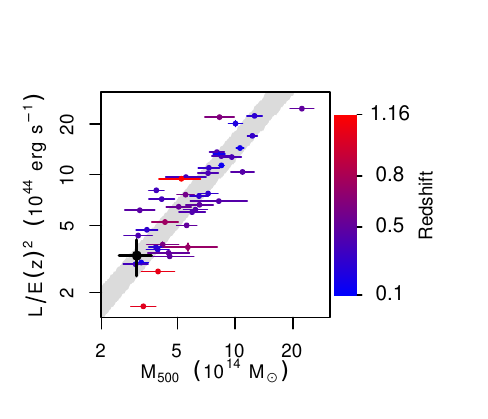}
  \includegraphics[scale=1.15,trim={10 17 20 33},clip]{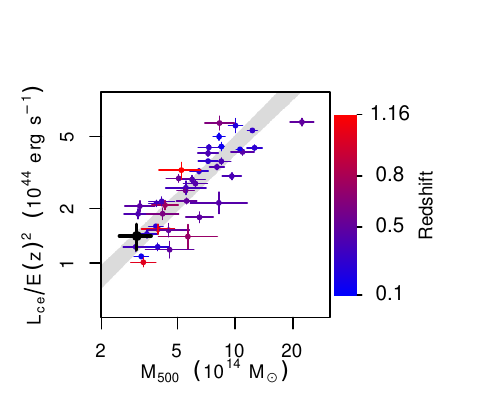}
  \caption{\small\protect\rule{0ex}{0ex}
    Mass, temperature, luminosity, and core-excised luminosity of ACT-CL~J0123 compared to that of the relaxed cluster samples from \cite{Mantz1509.01322}. Clusters are colored in order of increasing redshift (blue $\rightarrow$ red), with ACT-CL~J0123 in black. The grey shading is the 1$\sigma$ predictive range (including intrinsic scatter) surrounding a best fit powerlaw to the data in \cite{Mantz1509.01322} out to $z=1.06$. The reddest point (SPT-CL~J2215$-$3537 at $z=1.16$; \citealt{Stueber2601.14425}) is not included in this fit. ACT-CL~J0123 is consistent with the existing scaling relations.
  }
  \label{fig:scaling}
\end{figure}

\begin{table*}[ht]
\begin{center}
\begin{tabular}{ccccccc}
\hline
$z_\mathrm{spec}$ & r$_{500}$ (kpc) &M$_{500}$ (10$^{14}$ M$_\odot$)& M$_\mathrm{gas}$ (10$^{13}$ M$_\odot$) & kT (keV) & L$_{0.1-2.4}$ (10$^{45}$ ergs s$^{-1}$) & Z (Solar)  \\ \hline
      $1.50\pm0.03$ &  $590\pm40$& $3.1\pm0.6$ & $3.8\pm0.7$ & $7.3\pm1.1$ & $1.8 \pm 0.4$ & $0.43^{+0.46}_{-0.25}$ \\ \hline 
\end{tabular}
\caption{Global properties of ACT-CL~J0123. A redshift of $z=1.5$ has been assumed in the calculation of each property. Gas mass and $r_{500}$ (and consequently $M_{500}$) are simultaneously constrained using the 3-D deprojected radial profiles (Sections~\ref{sec:deprojection}/\ref{sec:mass}), while luminosity has been calculated in projection within $r_{500}$.}
\label{tab:bulk}
\end{center}
\end{table*}

ACT-CL~J0123 follows this general picture well. Figure~\ref{fig:thermo}, Figure~\ref{fig:dens_unscal}, and Figure~\ref{fig:scaling} each show a comparison of the properties of ACT-CL~J0123 against those of the relaxed galaxy cluster sample of \citep{Mantz1509.01322}, spanning the redshift range $0\lesssim z \lesssim 1.063$. Fig~\ref{fig:thermo} presents the radial profiles of ICM electron density, temperature, pseudo-entropy, and cooling time both unscaled and scaled self similarly. For these profiles, the ACT-CL~J0123 measurements from Chandra and XMM have been combined, using Chandra for the inner 100 kpc where we require high spatial resolution to resolve the cool core, and XMM for the remaining bins where we lack the statistics from Chandra to make precise measurements. The scaled measurements agree well with the full relaxed sample at large radius, providing evidence for self-similar evolution at intermediate-to-large radii in the dynamically relaxed cluster population out to $z=1.5$. We note, however, that the scaled density of the innermost bin is lower than the mean profile. Fig~\ref{fig:dens_unscal} presents the central density of ACT-CL~J0123 compared to the relaxed sample without scaling, where we find better agreement. This result extends the findings of \cite{McDonald1702.05094} which suggests that the physical state of the centers of cool core clusters are maintained to be approximately constant by AGN feedback since high redshift.

Figure~\ref{fig:scaling} presents the mass, temperature, and luminosity of ACT-CL~J0123 compared to the galaxy cluster scaling relations measured in \cite{Mantz1509.01322} in addition to the relaxed cluster SPT-CL~J2215$-$3537 at $z=1.16$ \citep{Stueber2601.14425}. Because ACT-CL~J0123 does not have deep enough data to constrain its mass directly, the value of $M_{500}$ reported here is the gas mass $M_\mathrm{gas}$ divided by the fiducial value of $f_\mathrm{gas}=0.125$ at $r_{500}$ \citep{Mantz1509.01322}. Luminosity is calculated in the 0.1--2.4 keV band, and we report both the total luminosity of the cluster, as well as the center-excised luminosity. The center-excised luminosity, $L_\mathrm{ce}$, excludes radii $r<0.15r_{500}$ in projection, and typically provides a lower-scatter mass proxy (bottom panel of Fig~\ref{fig:scaling}; see \citealt{Mantz1509.01322, Mantz1705.09329} for further details). Measurements of the temperature similarly excluded constraints from the central radii where the cool core would have a strong effect, though this measurement comes solely from XMM, which does not resolve the central drop in temperature in any case. As with $M_{500}$ (Section~\ref{sec:mass}), the uncertainty on these integrated quantities incorporates the uncertainty on $r_{500}$ itself. Overall, ACT-CL~J0123 is highly consistent with the existing population of relaxed clusters and supports previous suggestions that the bulk of clusters generally evolve self-similarly with the exception of the inner density profiles of relaxed/cool core systems. These global properties are reported in Table~\ref{tab:bulk}


\section{Summary and Conclusions}\label{sec:conclusions}
We report the identification of ACT-CL~J0123.5$-$0428 as the highest-redshift dynamically relaxed, cool core galaxy cluster discovered to date. Our primary findings can be summarized as follows: 
\begin{itemize}
    \item Our Chandra X-ray imaging analysis of ACT-CL~J0123.5$-$0428 shows it to be morphologically relaxed according to the SPA criteria of \cite{Mantz1502.06020}, with symmetry, peakiness, and alignment metrics of $s=1.3\pm0.2$, $p=-0.7\pm0.2$, and $a=1.3\pm0.2$, respectively.
    \item An XMM spectral analysis provides the first spectroscopic confirmation of ACT-CL~J0123 at $z=1.50\pm0.03$, making it the highest redshift dynamically relaxed cool core cluster identified to date, providing a substantial increase in redshift compared to the existing samples of relaxed systems ($z\leq 1.16$; \citealt{Mantz2111.09343}).
    \item The Chandra and XMM data confirm ACT-CL~J0123 to be one of the hottest ($kT=7.3\pm1.1$ keV), most X-ray luminous ($L_\mathrm{X}=1.8\pm0.4 \times 10^{45}$ ergs/s) and massive clusters ($M_{500} = 3.1\pm 0.6 \times 10^{14} \  \mathrm{M}_\odot$) known at high redshifts.   
    \item A Chandra spectral analysis of the inner 40 kpc of this system reveals a strong cool core ($1.8\pm0.6$ keV) with a low cooling time (t$_\mathrm{cool}=280^{+150}_{-120}$ Myr). 

    \item A deprojected spectral analysis of the Chandra and XMM data is able to jointly resolve ICM thermodynamic properties both within r$_{2500}$ and out to r$_{500}$. The mean, emission-weighted ICM metallicity is constrained to be $Z/Z_\odot=0.43^{+0.46}_{-0.25}$. 
    \item Comparison of the observed properties of ACT-CL~J0123 to an existing sample of relaxed clusters \citep{Mantz1509.01322} provides further evidence that clusters evolve self-similarly at large scales, while the inner ICM densities of cool core clusters remain roughly constant across cosmic time.
    \item A photometric analysis of the BCG of ACT-CL~J0123 reveals an evolved stellar population with no evidence for recent star formation. The data are consistent with an instantaneous starburst that occurred $\sim 2.3$--$3.5$ Gyr after the Big Bang.
\end{itemize}

ACT-CL~J0123.5$-$0428 is a prime example of why both high spatial resolution and high sensitivity are needed for the study of relaxed clusters at high redshift. With deeper data, further work could be done to better resolve the structure of the cool core and the distribution of metallicity within the cluster as a whole. At this time, supplemental XMM data will be essentially required for the most complete studies of high-redshift clusters, as Chandra's soft sensitivity continues to decrease and observations of this type become even more expensive to obtain. Accordingly, we highlight this study as a pathfinder for the next generation of X-ray missions, where instruments with high spatial resolution and high sensitivity (e.g.\ AXIS and NewAthena) will complement each other to continue this work at $z>1.5$.

\begin{acknowledgements}
AMF is supported by the Future Investigators in NASA Earth and Space Science and Technology (FINESST) Program under award number 80NSSC23K1484. Support for this work was provided in part by the National Aeronautics and Space Administration through Chandra Award Number GO3-24111A issued by the Chandra X-ray Center, which is operated by the Smithsonian Astrophysical Observatory for and on behalf of the National Aeronautics Space Administration under contract NAS8-03060. Work supported in part the U.S. Department of Energy under contract number DE-AC02-76SF00515. This research has made use of data obtained from the Chandra Data Archive provided by the Chandra X-ray Center (CXC) in addition to observations obtained with XMM-Newton, an ESA science mission
with instruments and contributions directly funded by
ESA Member States and NASA.
\end{acknowledgements}

\facilities{CXO, XMM}

\software{
  ACX (\citealt{Smith2012AN....333..301S, Smith1406.2037}; \url{http://atomdb.org/CX}),
  Astropy \citep{astropy1304.002, Astropy-Collaboration1307.6212, Astropy-Collaboration1801.02634, Astropy-Collaboration2206.14220},
  CIAO \citep{Fruscione2006SPIE.6270E..1VF, ciao1311.006},
  ESAS (\url{https://heasarc.gsfc.nasa.gov/docs/xmm/esas}),
  HEASOFT \citep{heasoft1408.004},
  LMC \citep{lmc1706.005},
  MARX \citep{Davis2012SPIE.8443E..1AD, marx1302.001},
  SAS \citep{sas1404.004},
  SXRBG \citep{sxrbg1904.001},
  XSPEC \citep{Arnaud1996ASPC..101...17A, xspec9910.005},
  Prospector \citep{Johnson2012.01426},
  extinction \citep{barbary_2016_804967}
}


\bibliography{references}{}

@ARTICLE{Scott1112.3769,
       author = {{Kay}, Scott T. and {Peel}, Michael W. and {Short}, C.~J. and {Thomas}, Peter A. and {Young}, Owain E. and {Battye}, Richard A. and {Liddle}, Andrew R. and {Pearce}, Frazer R.},
        title = "{Sunyaev-Zel'dovich clusters in Millennium gas simulations}",
      journal = {\mnras},
         year = 2012,
        month = may,
       volume = {422},
       number = {3},
        pages = {1999-2023},
          doi = {10.1111/j.1365-2966.2012.20623.x},
archivePrefix = {arXiv},
       eprint = {1112.3769},
 primaryClass = {astro-ph.CO},
       adsurl = {https://ui.adsabs.harvard.edu/abs/2012MNRAS.422.1999K}
}

@ARTICLE{Pop2205.11528,
       author = {{Pop}, Ana-Roxana and {Hernquist}, Lars and {Nagai}, Daisuke and {Kannan}, Rahul and {Weinberger}, Rainer and {Springel}, Volker and {Vogelsberger}, Mark and {Nelson}, Dylan and {Pakmor}, R{\"u}diger and {Pillepich}, Annalisa and {Torrey}, Paul},
        title = "{Sunyaev-Zel'dovich effect and X-ray scaling relations of galaxies, groups and clusters in the IllustrisTNG simulations}",
      journal = {arXiv e-prints},
         year = 2022,
        month = may,
          eid = {arXiv:2205.11528},
        pages = {arXiv:2205.11528},
          doi = {10.48550/arXiv.2205.11528},
archivePrefix = {arXiv},
       eprint = {2205.11528},
 primaryClass = {astro-ph.GA},
       adsurl = {https://ui.adsabs.harvard.edu/abs/2022arXiv220511528P}
}

@ARTICLE{Rasia2505.21624,
       author = {{Rasia}, Elena and {Tripodi}, Roberta and {Borgani}, Stefano and {Biffi}, Veronica and {Avestruz}, Camille and {Cui}, Weiguang and {De Petris}, Marco and {Dolag}, Klaus and {Eckert}, Dominique and {Ettori}, Stefano and {Gaspari}, Massimo},
        title = "{The Three Hundred Project: Modeling baryon and hot-gas fraction evolution in simulated clusters}",
      journal = {\aap},
         year = 2025,
        month = oct,
       volume = {702},
          eid = {A182},
        pages = {A182},
          doi = {10.1051/0004-6361/202554283},
archivePrefix = {arXiv},
       eprint = {2505.21624},
 primaryClass = {astro-ph.CO},
       adsurl = {https://ui.adsabs.harvard.edu/abs/2025A&A...702A.182R}
}

@ARTICLE{Stueber2601.14425,
       author = {{Stueber}, Haley R. and {Mantz}, Adam B. and {Allen}, Steven W. and {Flores}, Anthony M. and {Morris}, R. Glenn and {Pan}, Abigail Y. and {Somboonpanyakul}, Taweewat and {Bleem}, Lindsey E. and {Calzadilla}, Michael and {Floyd}, Benjamin and {Hlavacek-Larrondo}, Julie and {McDonald}, Michael and {Sarkar}, Arnab},
        title = "{Deep Chandra Observations of the z = 1.16 Relaxed, Cool-core Galaxy Cluster SPT-CL J2215-3537}",
      journal = {arXiv e-prints},
         year = 2026,
        month = jan,
          eid = {arXiv:2601.14425},
        pages = {arXiv:2601.14425},
          doi = {10.48550/arXiv.2601.14425},
archivePrefix = {arXiv},
       eprint = {2601.14425},
 primaryClass = {astro-ph.CO},
       adsurl = {https://ui.adsabs.harvard.edu/abs/2026arXiv260114425S}
}

@ARTICLE{Mantz2512.05405,
       author = {{Mantz}, Adam B. and {Flores}, Anthony M. and {Somboonpanyakul}, Taweewat and {Allen}, Steven W. and {Morris}, R. Glenn and {Pan}, Abigail Y. and {Stueber}, Haley R.},
        title = "{Ruminations Upon the Modeling of X-ray Foregrounds, Backgrounds and Faint Sources}",
      journal = {arXiv e-prints},
         year = 2025,
        month = dec,
          eid = {arXiv:2512.05405},
        pages = {arXiv:2512.05405},
          doi = {10.48550/arXiv.2512.05405},
archivePrefix = {arXiv},
       eprint = {2512.05405},
 primaryClass = {astro-ph.IM},
       adsurl = {https://ui.adsabs.harvard.edu/abs/2025arXiv251205405M}
}

@software{barbary_2016_804967,
author = {Barbary, Kyle},
title = {extinction v0.3.0},
month = dec,
year = 2016,
publisher = {Zenodo},
doi = {10.5281/zenodo.804967},
url = {https://doi.org/10.5281/zenodo.804967},
}

@ARTICLE{Aljamal2507.05176,
       author = {{Aljamal}, Eddie and {Evrard}, August E. and {Farahi}, Arya and {Pillepich}, Annalisa and {Nelson}, Dylan and {Schaye}, Joop and {Schaller}, Matthieu and {Braspenning}, Joey},
        title = "{Mass proxy quality of massive halo properties in the ILLUSTRISTNG and FLAMINGO simulations: I. Hot gas}",
      journal = {\mnras},
         year = 2025,
        month = nov,
       volume = {544},
       number = {1},
        pages = {67-94},
          doi = {10.1093/mnras/staf1665},
archivePrefix = {arXiv},
       eprint = {2507.05176},
 primaryClass = {astro-ph.GA},
       adsurl = {https://ui.adsabs.harvard.edu/abs/2025MNRAS.544...67A}
}

@ARTICLE{Ehlert2204.01765,
       author = {{Ehlert}, K. and {Weinberger}, R. and {Pfrommer}, C. and {Pakmor}, R. and {Springel}, V.},
        title = "{Self-regulated AGN feedback of light jets in cool-core galaxy clusters}",
      journal = {\mnras},
         year = 2023,
        month = jan,
       volume = {518},
       number = {3},
        pages = {4622-4645},
          doi = {10.1093/mnras/stac2860},
archivePrefix = {arXiv},
       eprint = {2204.01765},
 primaryClass = {astro-ph.GA},
       adsurl = {https://ui.adsabs.harvard.edu/abs/2023MNRAS.518.4622E}
}

@ARTICLE{Yang1605.01725,
       author = {{Yang}, H.-Y. Karen and {Reynolds}, Christopher S.},
        title = "{How AGN Jets Heat the Intracluster Medium{\textemdash}Insights from Hydrodynamic Simulations}",
      journal = {\apj},
         year = 2016,
        month = oct,
       volume = {829},
       number = {2},
          eid = {90},
        pages = {90},
          doi = {10.3847/0004-637X/829/2/90},
archivePrefix = {arXiv},
       eprint = {1605.01725},
 primaryClass = {astro-ph.HE},
       adsurl = {https://ui.adsabs.harvard.edu/abs/2016ApJ...829...90Y}
}

@ARTICLE{Li1611.05455,
       author = {{Li}, Yuan and {Ruszkowski}, Mateusz and {Bryan}, Greg L.},
        title = "{AGN Heating in Simulated Cool-core Clusters}",
      journal = {\apj},
         year = 2017,
        month = oct,
       volume = {847},
       number = {2},
          eid = {106},
        pages = {106},
          doi = {10.3847/1538-4357/aa88c1},
archivePrefix = {arXiv},
       eprint = {1611.05455},
 primaryClass = {astro-ph.GA},
       adsurl = {https://ui.adsabs.harvard.edu/abs/2017ApJ...847..106L}
}

@ARTICLE{Li1503.02660,
       author = {{Li}, Yuan and {Bryan}, Greg L. and {Ruszkowski}, Mateusz and {Voit}, G. Mark and {O'Shea}, Brian W. and {Donahue}, Megan},
        title = "{Cooling, AGN Feedback, and Star Formation in Simulated Cool-core Galaxy Clusters}",
      journal = {\apj},
         year = 2015,
        month = oct,
       volume = {811},
       number = {2},
          eid = {73},
        pages = {73},
          doi = {10.1088/0004-637X/811/2/73},
archivePrefix = {arXiv},
       eprint = {1503.02660},
 primaryClass = {astro-ph.GA},
       adsurl = {https://ui.adsabs.harvard.edu/abs/2015ApJ...811...73L}
}

@ARTICLE{Kriek1308.1099,
       author = {{Kriek}, Mariska and {Conroy}, Charlie},
        title = "{The Dust Attenuation Law in Distant Galaxies: Evidence for Variation with Spectral Type}",
      journal = {\apjl},
         year = 2013,
        month = sep,
       volume = {775},
       number = {1},
          eid = {L16},
        pages = {L16},
          doi = {10.1088/2041-8205/775/1/L16},
archivePrefix = {arXiv},
       eprint = {1308.1099},
 primaryClass = {astro-ph.CO},
       adsurl = {https://ui.adsabs.harvard.edu/abs/2013ApJ...775L..16K}
}

@software{sxrbg1904.001,
	adsurl = {https://ui.adsabs.harvard.edu/abs/2019ascl.soft04001S},
	author = {{Sabol}, Edward J. and {Snowden}, Steven L.},
	eid = {ascl:1904.001},
	howpublished = {Astrophysics Source Code Library, record ascl:1904.001},
	month = apr,
	title = {{sxrbg: ROSAT X-Ray Background Tool}},
	year = 2019}

@software{xspec9910.005,
	adsurl = {https://ui.adsabs.harvard.edu/abs/1999ascl.soft10005A},
	author = {{Arnaud}, Keith and {Dorman}, Ben and {Gordon}, Craig},
	eid = {ascl:9910.005},
	howpublished = {Astrophysics Source Code Library, record ascl:9910.005},
	month = oct,
	title = {{XSPEC: An X-ray spectral fitting package}},
	year = 1999}

@software{astropy1304.002,
	adsurl = {https://ui.adsabs.harvard.edu/abs/2013ascl.soft04002G},
	author = {{Greenfield}, Perry and {Robitaille}, Thomas and {Tollerud}, Erik and {Aldcroft}, Tom and {Barbary}, Kyle and {Barrett}, Paul and {Bray}, Erik and {Crighton}, Neil and {Conley}, Alex and {Conseil}, Simon and {Davis}, Matt and {Deil}, Christoph and {Dencheva}, Nadia and {Droettboom}, Michael and {Ferguson}, Henry and {Ginsburg}, Adam and {Grollier}, Fr{\'e}d{\'e}ric and {Moritz G{\"u}nther}, Hans and {Hanley}, Chris and {Hsu}, J.~C. and {Kerzendorf}, Wolfgang and {Kramer}, Roban and {Lian Lim}, Pey and {Muna}, Demitri and {Nair}, Prasanth and {Price-Whelan}, Adrian and {Shiga}, David and {Singer}, Leo and {Taylor}, James and {Turner}, James and {Woillez}, Julien and {Zabalza}, Victor},
	eid = {ascl:1304.002},
	howpublished = {Astrophysics Source Code Library, record ascl:1304.002},
	month = apr,
	title = {{Astropy: Community Python library for astronomy}},
	year = 2013}

@software{sas1404.004,
	adsurl = {https://ui.adsabs.harvard.edu/abs/2014ascl.soft04004S},
	author = {{SAS development team}},
	eid = {ascl:1404.004},
	howpublished = {Astrophysics Source Code Library, record ascl:1404.004},
	month = apr,
	title = {{SAS: Science Analysis System for XMM-Newton observatory}},
	year = 2014}

@inproceedings{Davis2012SPIE.8443E..1AD,
	adsurl = {https://ui.adsabs.harvard.edu/abs/2012SPIE.8443E..1AD},
	author = {{Davis}, John E. and {Bautz}, Marshall W. and {Dewey}, Daniel and {Heilmann}, Ralf K. and {Houck}, John C. and {Huenemoerder}, David P. and {Marshall}, Herman L. and {Nowak}, Michael A. and {Schattenburg}, Mark L. and {Schulz}, Norbert S. and {Smith}, Randall K.},
	booktitle = {Space Telescopes and Instrumentation 2012: Ultraviolet to Gamma Ray},
	doi = {10.1117/12.926937},
	editor = {{Takahashi}, Tadayuki and {Murray}, Stephen S. and {den Herder}, Jan-Willem A.},
	eid = {84431A},
	month = sep,
	pages = {84431A},
	series = {Society of Photo-Optical Instrumentation Engineers (SPIE) Conference Series},
	title = {{Raytracing with MARX: x-ray observatory design, calibration, and support}},
	volume = {8443},
	year = 2012}

@software{marx1302.001,
	adsurl = {https://ui.adsabs.harvard.edu/abs/2013ascl.soft02001W},
	author = {{Wise}, Michael W. and {Davis}, John E. and {Huenemoerder}, David P. and {Houck}, John C. and {Dewey}, Dan},
	eid = {ascl:1302.001},
	howpublished = {Astrophysics Source Code Library, record ascl:1302.001},
	month = feb,
	title = {{MARX: Model of AXAF Response to X-rays}},
	year = 2013}

@software{lmc1706.005,
	adsurl = {https://ui.adsabs.harvard.edu/abs/2017ascl.soft06005M},
	author = {{Mantz}, Adam B.},
	eid = {ascl:1706.005},
	howpublished = {Astrophysics Source Code Library, record ascl:1706.005},
	month = jun,
	title = {{LMC: Logarithmantic Monte Carlo}},
	year = 2017}

@software{heasoft1408.004,
	adsurl = {https://ui.adsabs.harvard.edu/abs/2014ascl.soft08004N},
	author = {{NASA High Energy Astrophysics Science Archive Research Center (Heasarc)}},
	eid = {ascl:1408.004},
	howpublished = {Astrophysics Source Code Library, record ascl:1408.004},
	month = aug,
	title = {{HEAsoft: Unified Release of FTOOLS and XANADU}},
	year = 2014}

@inproceedings{Fruscione2006SPIE.6270E..1VF,
	adsurl = {https://ui.adsabs.harvard.edu/abs/2006SPIE.6270E..1VF},
	author = {{Fruscione}, Antonella and {McDowell}, Jonathan C. and {Allen}, Glenn E. and {Brickhouse}, Nancy S. and {Burke}, Douglas J. and {Davis}, John E. and {Durham}, Nick and {Elvis}, Martin and {Galle}, Elizabeth C. and {Harris}, Daniel E. and {Huenemoerder}, David P. and {Houck}, John C. and {Ishibashi}, Bish and {Karovska}, Margarita and {Nicastro}, Fabrizio and {Noble}, Michael S. and {Nowak}, Michael A. and {Primini}, Frank A. and {Siemiginowska}, Aneta and {Smith}, Randall K. and {Wise}, Michael},
	booktitle = {Observatory Operations: Strategies, Processes, and Systems},
	doi = {10.1117/12.671760},
	editor = {{Silva}, David R. and {Doxsey}, Rodger E.},
	eid = {62701V},
	month = jun,
	pages = {62701V},
	series = {Society of Photo-Optical Instrumentation Engineers (SPIE) Conference Series},
	title = {{CIAO: Chandra's data analysis system}},
	volume = {6270},
	year = 2006}

@software{ciao1311.006,
	adsurl = {https://ui.adsabs.harvard.edu/abs/2013ascl.soft11006C},
	author = {{CIAO Development Team}},
	eid = {ascl:1311.006},
	howpublished = {Astrophysics Source Code Library, record ascl:1311.006},
	month = nov,
	title = {{CIAO: Chandra Interactive Analysis of Observations}},
	year = 2013}

@article{Astropy-Collaboration2206.14220,
	adsurl = {https://ui.adsabs.harvard.edu/abs/2022ApJ...935..167A},
	archiveprefix = {arXiv},
	author = {{Astropy Collaboration} and {Price-Whelan}, Adrian M. and {Lim}, Pey Lian and {Earl}, Nicholas and {Starkman}, Nathaniel and {Bradley}, Larry and {Shupe}, David L. and {Patil}, Aarya A. and {Corrales}, Lia and {Brasseur}, C.~E. and {N{\"o}the}, Maximilian and {Donath}, Axel and {Tollerud}, Erik and {Morris}, Brett M. and {Ginsburg}, Adam and {Vaher}, Eero and {Weaver}, Benjamin A. and {Tocknell}, James and {Jamieson}, William and {van Kerkwijk}, Marten H. and {Robitaille}, Thomas P. and {Merry}, Bruce and {Bachetti}, Matteo and {G{\"u}nther}, H. Moritz and {Aldcroft}, Thomas L. and {Alvarado-Montes}, Jaime A. and {Archibald}, Anne M. and {B{\'o}di}, Attila and {Bapat}, Shreyas and {Barentsen}, Geert and {Baz{\'a}n}, Juanjo and {Biswas}, Manish and {Boquien}, M{\'e}d{\'e}ric and {Burke}, D.~J. and {Cara}, Daria and {Cara}, Mihai and {Conroy}, Kyle E. and {Conseil}, Simon and {Craig}, Matthew W. and {Cross}, Robert M. and {Cruz}, Kelle L. and {D'Eugenio}, Francesco and {Dencheva}, Nadia and {Devillepoix}, Hadrien A.~R. and {Dietrich}, J{\"o}rg P. and {Eigenbrot}, Arthur Davis and {Erben}, Thomas and {Ferreira}, Leonardo and {Foreman-Mackey}, Daniel and {Fox}, Ryan and {Freij}, Nabil and {Garg}, Suyog and {Geda}, Robel and {Glattly}, Lauren and {Gondhalekar}, Yash and {Gordon}, Karl D. and {Grant}, David and {Greenfield}, Perry and {Groener}, Austen M. and {Guest}, Steve and {Gurovich}, Sebastian and {Handberg}, Rasmus and {Hart}, Akeem and {Hatfield-Dodds}, Zac and {Homeier}, Derek and {Hosseinzadeh}, Griffin and {Jenness}, Tim and {Jones}, Craig K. and {Joseph}, Prajwel and {Kalmbach}, J. Bryce and {Karamehmetoglu}, Emir and {Ka{\l}uszy{\'n}ski}, Miko{\l}aj and {Kelley}, Michael S.~P. and {Kern}, Nicholas and {Kerzendorf}, Wolfgang E. and {Koch}, Eric W. and {Kulumani}, Shankar and {Lee}, Antony and {Ly}, Chun and {Ma}, Zhiyuan and {MacBride}, Conor and {Maljaars}, Jakob M. and {Muna}, Demitri and {Murphy}, N.~A. and {Norman}, Henrik and {O'Steen}, Richard and {Oman}, Kyle A. and {Pacifici}, Camilla and {Pascual}, Sergio and {Pascual-Granado}, J. and {Patil}, Rohit R. and {Perren}, Gabriel I. and {Pickering}, Timothy E. and {Rastogi}, Tanuj and {Roulston}, Benjamin R. and {Ryan}, Daniel F. and {Rykoff}, Eli S. and {Sabater}, Jose and {Sakurikar}, Parikshit and {Salgado}, Jes{\'u}s and {Sanghi}, Aniket and {Saunders}, Nicholas and {Savchenko}, Volodymyr and {Schwardt}, Ludwig and {Seifert-Eckert}, Michael and {Shih}, Albert Y. and {Jain}, Anany Shrey and {Shukla}, Gyanendra and {Sick}, Jonathan and {Simpson}, Chris and {Singanamalla}, Sudheesh and {Singer}, Leo P. and {Singhal}, Jaladh and {Sinha}, Manodeep and {Sip{\H{o}}cz}, Brigitta M. and {Spitler}, Lee R. and {Stansby}, David and {Streicher}, Ole and {{\v{S}}umak}, Jani and {Swinbank}, John D. and {Taranu}, Dan S. and {Tewary}, Nikita and {Tremblay}, Grant R. and {de Val-Borro}, Miguel and {Van Kooten}, Samuel J. and {Vasovi{\'c}}, Zlatan and {Verma}, Shresth and {de Miranda Cardoso}, Jos{\'e} Vin{\'\i}cius and {Williams}, Peter K.~G. and {Wilson}, Tom J. and {Winkel}, Benjamin and {Wood-Vasey}, W.~M. and {Xue}, Rui and {Yoachim}, Peter and {Zhang}, Chen and {Zonca}, Andrea and {Astropy Project Contributors}},
	doi = {10.3847/1538-4357/ac7c74},
	eid = {167},
	eprint = {2206.14220},
	journal = {\apj},
	month = aug,
	number = {2},
	pages = {167},
	primaryclass = {astro-ph.IM},
	title = {{The Astropy Project: Sustaining and Growing a Community-oriented Open-source Project and the Latest Major Release (v5.0) of the Core Package}},
	volume = {935},
	year = 2022}

@article{Astropy-Collaboration1801.02634,
	adsurl = {https://ui.adsabs.harvard.edu/abs/2018AJ....156..123A},
	archiveprefix = {arXiv},
	author = {{Astropy Collaboration} and {Price-Whelan}, A.~M. and {Sip{\H{o}}cz}, B.~M. and {G{\"u}nther}, H.~M. and {Lim}, P.~L. and {Crawford}, S.~M. and {Conseil}, S. and {Shupe}, D.~L. and {Craig}, M.~W. and {Dencheva}, N. and {Ginsburg}, A. and {VanderPlas}, J.~T. and {Bradley}, L.~D. and {P{\'e}rez-Su{\'a}rez}, D. and {de Val-Borro}, M. and {Aldcroft}, T.~L. and {Cruz}, K.~L. and {Robitaille}, T.~P. and {Tollerud}, E.~J. and {Ardelean}, C. and {Babej}, T. and {Bach}, Y.~P. and {Bachetti}, M. and {Bakanov}, A.~V. and {Bamford}, S.~P. and {Barentsen}, G. and {Barmby}, P. and {Baumbach}, A. and {Berry}, K.~L. and {Biscani}, F. and {Boquien}, M. and {Bostroem}, K.~A. and {Bouma}, L.~G. and {Brammer}, G.~B. and {Bray}, E.~M. and {Breytenbach}, H. and {Buddelmeijer}, H. and {Burke}, D.~J. and {Calderone}, G. and {Cano Rodr{\'\i}guez}, J.~L. and {Cara}, M. and {Cardoso}, J.~V.~M. and {Cheedella}, S. and {Copin}, Y. and {Corrales}, L. and {Crichton}, D. and {D'Avella}, D. and {Deil}, C. and {Depagne}, {\'E}. and {Dietrich}, J.~P. and {Donath}, A. and {Droettboom}, M. and {Earl}, N. and {Erben}, T. and {Fabbro}, S. and {Ferreira}, L.~A. and {Finethy}, T. and {Fox}, R.~T. and {Garrison}, L.~H. and {Gibbons}, S.~L.~J. and {Goldstein}, D.~A. and {Gommers}, R. and {Greco}, J.~P. and {Greenfield}, P. and {Groener}, A.~M. and {Grollier}, F. and {Hagen}, A. and {Hirst}, P. and {Homeier}, D. and {Horton}, A.~J. and {Hosseinzadeh}, G. and {Hu}, L. and {Hunkeler}, J.~S. and {Ivezi{\'c}}, {\v{Z}}. and {Jain}, A. and {Jenness}, T. and {Kanarek}, G. and {Kendrew}, S. and {Kern}, N.~S. and {Kerzendorf}, W.~E. and {Khvalko}, A. and {King}, J. and {Kirkby}, D. and {Kulkarni}, A.~M. and {Kumar}, A. and {Lee}, A. and {Lenz}, D. and {Littlefair}, S.~P. and {Ma}, Z. and {Macleod}, D.~M. and {Mastropietro}, M. and {McCully}, C. and {Montagnac}, S. and {Morris}, B.~M. and {Mueller}, M. and {Mumford}, S.~J. and {Muna}, D. and {Murphy}, N.~A. and {Nelson}, S. and {Nguyen}, G.~H. and {Ninan}, J.~P. and {N{\"o}the}, M. and {Ogaz}, S. and {Oh}, S. and {Parejko}, J.~K. and {Parley}, N. and {Pascual}, S. and {Patil}, R. and {Patil}, A.~A. and {Plunkett}, A.~L. and {Prochaska}, J.~X. and {Rastogi}, T. and {Reddy Janga}, V. and {Sabater}, J. and {Sakurikar}, P. and {Seifert}, M. and {Sherbert}, L.~E. and {Sherwood-Taylor}, H. and {Shih}, A.~Y. and {Sick}, J. and {Silbiger}, M.~T. and {Singanamalla}, S. and {Singer}, L.~P. and {Sladen}, P.~H. and {Sooley}, K.~A. and {Sornarajah}, S. and {Streicher}, O. and {Teuben}, P. and {Thomas}, S.~W. and {Tremblay}, G.~R. and {Turner}, J.~E.~H. and {Terr{\'o}n}, V. and {van Kerkwijk}, M.~H. and {de la Vega}, A. and {Watkins}, L.~L. and {Weaver}, B.~A. and {Whitmore}, J.~B. and {Woillez}, J. and {Zabalza}, V. and {Astropy Contributors}},
	doi = {10.3847/1538-3881/aabc4f},
	eid = {123},
	eprint = {1801.02634},
	journal = {\aj},
	month = sep,
	number = {3},
	pages = {123},
	primaryclass = {astro-ph.IM},
	title = {{The Astropy Project: Building an Open-science Project and Status of the v2.0 Core Package}},
	volume = {156},
	year = 2018}

@article{Astropy-Collaboration1307.6212,
	adsurl = {https://ui.adsabs.harvard.edu/abs/2013A&A...558A..33A},
	archiveprefix = {arXiv},
	author = {{Astropy Collaboration} and {Robitaille}, Thomas P. and {Tollerud}, Erik J. and {Greenfield}, Perry and {Droettboom}, Michael and {Bray}, Erik and {Aldcroft}, Tom and {Davis}, Matt and {Ginsburg}, Adam and {Price-Whelan}, Adrian M. and {Kerzendorf}, Wolfgang E. and {Conley}, Alexander and {Crighton}, Neil and {Barbary}, Kyle and {Muna}, Demitri and {Ferguson}, Henry and {Grollier}, Fr{\'e}d{\'e}ric and {Parikh}, Madhura M. and {Nair}, Prasanth H. and {Unther}, Hans M. and {Deil}, Christoph and {Woillez}, Julien and {Conseil}, Simon and {Kramer}, Roban and {Turner}, James E.~H. and {Singer}, Leo and {Fox}, Ryan and {Weaver}, Benjamin A. and {Zabalza}, Victor and {Edwards}, Zachary I. and {Azalee Bostroem}, K. and {Burke}, D.~J. and {Casey}, Andrew R. and {Crawford}, Steven M. and {Dencheva}, Nadia and {Ely}, Justin and {Jenness}, Tim and {Labrie}, Kathleen and {Lim}, Pey Lian and {Pierfederici}, Francesco and {Pontzen}, Andrew and {Ptak}, Andy and {Refsdal}, Brian and {Servillat}, Mathieu and {Streicher}, Ole},
	doi = {10.1051/0004-6361/201322068},
	eid = {A33},
	eprint = {1307.6212},
	journal = {\aap},
	month = oct,
	pages = {A33},
	primaryclass = {astro-ph.IM},
	title = {{Astropy: A community Python package for astronomy}},
	volume = {558},
	year = 2013}

@article{Smith2012AN....333..301S,
	adsurl = {https://ui.adsabs.harvard.edu/abs/2012AN....333..301S},
	author = {{Smith}, R.~K. and {Foster}, A.~R. and {Brickhouse}, N.~S.},
	doi = {10.1002/asna.201211673},
	journal = {Astronomische Nachrichten},
	month = apr,
	number = {4},
	pages = {301},
	title = {{Approximating the X-ray spectrum emitted from astrophysical charge exchange}},
	volume = {333},
	year = 2012}

@ARTICLE{Allen0002506,
       author = {{Allen}, S.~W.},
        title = "{The properties of cooling flows in X-ray luminous clusters of galaxies}",
      journal = {\mnras},
         year = 2000,
        month = jun,
       volume = {315},
       number = {2},
        pages = {269-295},
          doi = {10.1046/j.1365-8711.2000.03395.x},
archivePrefix = {arXiv},
       eprint = {astro-ph/0002506},
 primaryClass = {astro-ph},
       adsurl = {https://ui.adsabs.harvard.edu/abs/2000MNRAS.315..269A}
}

@ARTICLE{Ruppin2207.13351,
       author = {{Ruppin}, F. and {McDonald}, M. and {Hlavacek-Larrondo}, J. and {Bayliss}, M. and {Bleem}, L.~E. and {Calzadilla}, M. and {Edge}, A.~C. and {Filipovi{\'c}}, M.~D. and {Floyd}, B. and {Garmire}, G. and {Khullar}, G. and {Kim}, K.~J. and {Kraft}, R. and {Mahler}, G. and {Norris}, R.~P. and {O'Brien}, A. and {Reichardt}, C.~L. and {Somboonpanyakul}, T. and {Stark}, A.~A. and {Tothill}, N.},
        title = "{Redshift Evolution of the Feedback-Cooling Equilibrium in the Core of 48 SPT Galaxy Clusters: A Joint Chandra-SPT-ATCA Analysis}",
      journal = {\apj},
         year = 2023,
        month = may,
       volume = {948},
       number = {1},
          eid = {49},
        pages = {49},
          doi = {10.3847/1538-4357/acc38d},
archivePrefix = {arXiv},
       eprint = {2207.13351},
 primaryClass = {astro-ph.CO},
       adsurl = {https://ui.adsabs.harvard.edu/abs/2023ApJ...948...49R}
}

@ARTICLE{Martz2003.11104,
       author = {{Martz}, C.~G. and {McNamara}, B.~R. and {Nulsen}, P.~E.~J. and {Vantyghem}, A.~N. and {Gingras}, M. -J. and {Babyk}, Iu. V. and {Russell}, H.~R. and {Edge}, A.~C. and {McDonald}, M. and {Tamhane}, P.~D. and {Fabian}, A.~C. and {Hogan}, M.~T.},
        title = "{Thermally Unstable Cooling Stimulated by Uplift: The Spoiler Clusters}",
      journal = {\apj},
         year = 2020,
        month = jul,
       volume = {897},
       number = {1},
          eid = {57},
        pages = {57},
          doi = {10.3847/1538-4357/ab96cd},
archivePrefix = {arXiv},
       eprint = {2003.11104},
 primaryClass = {astro-ph.GA},
       adsurl = {https://ui.adsabs.harvard.edu/abs/2020ApJ...897...57M}
}

@ARTICLE{Edge0106225,
       author = {{Edge}, A.~C.},
        title = "{The detection of molecular gas in the central galaxies of cooling flow clusters}",
      journal = {\mnras},
         year = 2001,
        month = dec,
       volume = {328},
       number = {3},
        pages = {762-782},
          doi = {10.1046/j.1365-8711.2001.04802.x},
archivePrefix = {arXiv},
       eprint = {astro-ph/0106225},
 primaryClass = {astro-ph},
       adsurl = {https://ui.adsabs.harvard.edu/abs/2001MNRAS.328..762E}
}

@ARTICLE{Edge1005.1207,
       author = {{Edge}, A.~C. and {Oonk}, J.~B.~R. and {Mittal}, R. and {Allen}, S.~W. and {Baum}, S.~A. and {B{\"o}hringer}, H. and {Bregman}, J.~N. and {Bremer}, M.~N. and {Combes}, F. and {Crawford}, C.~S. and {Donahue}, M. and {Egami}, E. and {Fabian}, A.~C. and {Ferland}, G.~J. and {Hamer}, S.~L. and {Hatch}, N.~A. and {Jaffe}, W. and {Johnstone}, R.~M. and {McNamara}, B.~R. and {O'Dea}, C.~P. and {Popesso}, P. and {Quillen}, A.~C. and {Salom{\'e}}, P. and {Sarazin}, C.~L. and {Voit}, G.~M. and {Wilman}, R.~J. and {Wise}, M.~W.},
        title = "{Herschel observations of FIR emission lines in brightest cluster galaxies}",
      journal = {\aap},
         year = 2010,
        month = jul,
       volume = {518},
          eid = {L46},
        pages = {L46},
          doi = {10.1051/0004-6361/201014569},
archivePrefix = {arXiv},
       eprint = {1005.1207},
 primaryClass = {astro-ph.CO},
       adsurl = {https://ui.adsabs.harvard.edu/abs/2010A&A...518L..46E}
}

@ARTICLE{Olivares1902.09164,
       author = {{Olivares}, V. and {Salome}, P. and {Combes}, F. and {Hamer}, S. and {Guillard}, P. and {Lehnert}, M.~D. and {Polles}, F.~L. and {Beckmann}, R.~S. and {Dubois}, Y. and {Donahue}, M. and {Edge}, A. and {Fabian}, A.~C. and {McNamara}, B. and {Rose}, T. and {Russell}, H.~R. and {Tremblay}, G. and {Vantyghem}, A. and {Canning}, R.~E.~A. and {Ferland}, G. and {Godard}, B. and {Peirani}, S. and {Pineau des Forets}, G.},
        title = "{Ubiquitous cold and massive filaments in cool core clusters}",
      journal = {\aap},
         year = 2019,
        month = nov,
       volume = {631},
          eid = {A22},
        pages = {A22},
          doi = {10.1051/0004-6361/201935350},
archivePrefix = {arXiv},
       eprint = {1902.09164},
 primaryClass = {astro-ph.GA},
       adsurl = {https://ui.adsabs.harvard.edu/abs/2019A&A...631A..22O}
}

@ARTICLE{Kravtsov1205.5556,
       author = {{Kravtsov}, Andrey V. and {Borgani}, Stefano},
        title = "{Formation of Galaxy Clusters}",
      journal = {\araa},
         year = 2012,
        month = sep,
       volume = {50},
        pages = {353-409},
          doi = {10.1146/annurev-astro-081811-125502},
archivePrefix = {arXiv},
       eprint = {1205.5556},
 primaryClass = {astro-ph.CO},
       adsurl = {https://ui.adsabs.harvard.edu/abs/2012ARA&A..50..353K}
}

@INCOLLECTION{Allen2019book,
       author = {{Allen}, Steven W. and {Mantz}, Adam B.},
        title = "{Galaxy Cluster Cosmology}",
    booktitle = {The Chandra X-ray Observatory},
         year = 2019,
       editor = {{Wilkes}, Belinda and {Tucker}, Wallace},
        pages = {10-1},
          doi = {10.1088/2514-3433/ab43dcch10},
       adsurl = {https://ui.adsabs.harvard.edu/abs/2019cxro.book...10A}
}

@ARTICLE{Jimenez-Teja2305.10860,
       author = {{Jim{\'e}nez-Teja}, Y. and {Dupke}, R.~A. and {Lopes}, P.~A.~A. and {V{\'\i}lchez}, J.~M.},
        title = "{Dissecting the RELICS cluster SPT-CLJ0615-5746 through intracluster light: Confirmation of the multiple merging state of the cluster formation}",
      journal = {\aap},
         year = 2023,
        month = aug,
       volume = {676},
          eid = {A39},
        pages = {A39},
          doi = {10.1051/0004-6361/202346580},
archivePrefix = {arXiv},
       eprint = {2305.10860},
 primaryClass = {astro-ph.GA},
       adsurl = {https://ui.adsabs.harvard.edu/abs/2023A&A...676A..39J}
}

@ARTICLE{Bartalucci1610.01899,
       author = {{Bartalucci}, I. and {Arnaud}, M. and {Pratt}, G.~W. and {D{\'e}mocl{\`e}s}, J. and {van der Burg}, R.~F.~J. and {Mazzotta}, P.},
        title = "{Resolving galaxy cluster gas properties at z   1 with XMM-Newton and Chandra}",
      journal = {\aap},
         year = 2017,
        month = feb,
       volume = {598},
          eid = {A61},
        pages = {A61},
          doi = {10.1051/0004-6361/201629509},
archivePrefix = {arXiv},
       eprint = {1610.01899},
 primaryClass = {astro-ph.CO},
       adsurl = {https://ui.adsabs.harvard.edu/abs/2017A&A...598A..61B}
}

@ARTICLE{Calzadilla2311.00396,
       author = {{Calzadilla}, Michael S. and {McDonald}, Michael and {Benson}, Bradford A. and {Bleem}, Lindsey E. and {Croston}, Judith H. and {Donahue}, Megan and {Edge}, Alastair C. and {Floyd}, Benjamin and {Garmire}, Gordon P. and {Hlavacek-Larrondo}, Julie and {Huynh}, Minh T. and {Khullar}, Gourav and {Kraft}, Ralph P. and {McNamara}, Brian R. and {Noble}, Allison G. and {Romero}, Charles E. and {Ruppin}, Florian and {Somboonpanyakul}, Taweewat and {Voit}, G. Mark},
        title = "{The SPT-Chandra BCG Spectroscopic Survey. I. Evolution of the Entropy Threshold for ICM Cooling and AGN Feedback in Galaxy Clusters over the Last 10 Gyr}",
      journal = {\apj},
         year = 2024,
        month = dec,
       volume = {976},
       number = {2},
          eid = {169},
        pages = {169},
          doi = {10.3847/1538-4357/ad8916},
archivePrefix = {arXiv},
       eprint = {2311.00396},
 primaryClass = {astro-ph.GA},
       adsurl = {https://ui.adsabs.harvard.edu/abs/2024ApJ...976..169C}
}

@ARTICLE{Voit0806.0384,
       author = {{Voit}, G.~M. and {Cavagnolo}, K.~W. and {Donahue}, M. and {Rafferty}, D.~A. and {McNamara}, B.~R. and {Nulsen}, P.~E.~J.},
        title = "{Conduction and the Star Formation Threshold in Brightest Cluster Galaxies}",
      journal = {\apjl},
         year = 2008,
        month = jul,
       volume = {681},
       number = {1},
        pages = {L5},
          doi = {10.1086/590344},
archivePrefix = {arXiv},
       eprint = {0806.0384},
 primaryClass = {astro-ph},
       adsurl = {https://ui.adsabs.harvard.edu/abs/2008ApJ...681L...5V}
}

@ARTICLE{Cavagnolo0806.0382,
       author = {{Cavagnolo}, Kenneth W. and {Donahue}, Megan and {Voit}, G. Mark and {Sun}, Ming},
        title = "{An Entropy Threshold for Strong H{\ensuremath{\alpha}} and Radio Emission in the Cores of Galaxy Clusters}",
      journal = {\apjl},
         year = 2008,
        month = aug,
       volume = {683},
       number = {2},
        pages = {L107},
          doi = {10.1086/591665},
archivePrefix = {arXiv},
       eprint = {0806.0382},
 primaryClass = {astro-ph},
       adsurl = {https://ui.adsabs.harvard.edu/abs/2008ApJ...683L.107C}
}

@ARTICLE{Donahue1004.0529,
       author = {{Donahue}, Megan and {Bruch}, Seth and {Wang}, Emily and {Voit}, G. Mark and {Hicks}, Amalia K. and {Haarsma}, Deborah B. and {Croston}, Judith H. and {Pratt}, Gabriel W. and {Pierini}, Daniele and {O'Connell}, Robert W. and {B{\"o}hringer}, Hans},
        title = "{Star Formation and UV colors of the Brightest Cluster Galaxies in the Representative XMM-Newton Cluster Structure Survey}",
      journal = {\apj},
         year = 2010,
        month = jun,
       volume = {715},
       number = {2},
        pages = {881-896},
          doi = {10.1088/0004-637X/715/2/881},
archivePrefix = {arXiv},
       eprint = {1004.0529},
 primaryClass = {astro-ph.CO},
       adsurl = {https://ui.adsabs.harvard.edu/abs/2010ApJ...715..881D}
}

@ARTICLE{Foreman-Mackey1202.3665,
       author = {{Foreman-Mackey}, Daniel and {Hogg}, David W. and {Lang}, Dustin and {Goodman}, Jonathan},
        title = "{emcee: The MCMC Hammer}",
      journal = {\pasp},
         year = 2013,
        month = mar,
       volume = {125},
       number = {925},
        pages = {306},
          doi = {10.1086/670067},
archivePrefix = {arXiv},
       eprint = {1202.3665},
 primaryClass = {astro-ph.IM},
       adsurl = {https://ui.adsabs.harvard.edu/abs/2013PASP..125..306F}
}

@ARTICLE{Schlafly1012.4804,
       author = {{Schlafly}, Edward F. and {Finkbeiner}, Douglas P.},
        title = "{Measuring Reddening with Sloan Digital Sky Survey Stellar Spectra and Recalibrating SFD}",
      journal = {\apj},
         year = 2011,
        month = aug,
       volume = {737},
       number = {2},
          eid = {103},
        pages = {103},
          doi = {10.1088/0004-637X/737/2/103},
archivePrefix = {arXiv},
       eprint = {1012.4804},
 primaryClass = {astro-ph.GA},
       adsurl = {https://ui.adsabs.harvard.edu/abs/2011ApJ...737..103S}
}

@ARTICLE{Cardelli1989,
       author = {{Cardelli}, Jason A. and {Clayton}, Geoffrey C. and {Mathis}, John S.},
        title = "{The Relationship between Infrared, Optical, and Ultraviolet Extinction}",
      journal = {\apj},
         year = 1989,
        month = oct,
       volume = {345},
        pages = {245},
          doi = {10.1086/167900},
       adsurl = {https://ui.adsabs.harvard.edu/abs/1989ApJ...345..245C}
}

@ARTICLE{Kroupa0009005,
       author = {{Kroupa}, Pavel},
        title = "{On the variation of the initial mass function}",
      journal = {\mnras},
         year = 2001,
        month = apr,
       volume = {322},
       number = {2},
        pages = {231-246},
          doi = {10.1046/j.1365-8711.2001.04022.x},
archivePrefix = {arXiv},
       eprint = {astro-ph/0009005},
 primaryClass = {astro-ph},
       adsurl = {https://ui.adsabs.harvard.edu/abs/2001MNRAS.322..231K}
}

@ARTICLE{Conroy0809.4261,
       author = {{Conroy}, Charlie and {Gunn}, James E. and {White}, Martin},
        title = "{The Propagation of Uncertainties in Stellar Population Synthesis Modeling. I. The Relevance of Uncertain Aspects of Stellar Evolution and the Initial Mass Function to the Derived Physical Properties of Galaxies}",
      journal = {\apj},
         year = 2009,
        month = jul,
       volume = {699},
       number = {1},
        pages = {486-506},
          doi = {10.1088/0004-637X/699/1/486},
archivePrefix = {arXiv},
       eprint = {0809.4261},
 primaryClass = {astro-ph},
       adsurl = {https://ui.adsabs.harvard.edu/abs/2009ApJ...699..486C}
}

@ARTICLE{Conroy0911.3151,
       author = {{Conroy}, Charlie and {Gunn}, James E.},
        title = "{The Propagation of Uncertainties in Stellar Population Synthesis Modeling. III. Model Calibration, Comparison, and Evaluation}",
      journal = {\apj},
         year = 2010,
        month = apr,
       volume = {712},
       number = {2},
        pages = {833-857},
          doi = {10.1088/0004-637X/712/2/833},
archivePrefix = {arXiv},
       eprint = {0911.3151},
 primaryClass = {astro-ph.CO},
       adsurl = {https://ui.adsabs.harvard.edu/abs/2010ApJ...712..833C}
}

@ARTICLE{Johnson2012.01426,
    author = {{Johnson}, Benjamin D. and {Leja}, Joel and {Conroy}, Charlie and {Speagle}, Joshua S.},
        title = "{Stellar Population Inference with Prospector}",
    journal = {\apjs},
        year = 2021,
        month = jun,
    volume = {254},
    number = {2},
        eid = {22},
        pages = {22},
        doi = {10.3847/1538-4365/abef67},
archivePrefix = {arXiv},
    eprint = {2012.01426},
primaryClass = {astro-ph.GA},
    adsurl = {https://ui.adsabs.harvard.edu/abs/2021ApJS..254...22J}
}

@ARTICLE{Eisenhardt1908.08902,
       author = {{Eisenhardt}, Peter R.~M. and {Marocco}, Federico and {Fowler}, John W. and {Meisner}, Aaron M. and {Kirkpatrick}, J. Davy and {Garcia}, Nelson and {Jarrett}, Thomas H. and {Koontz}, Renata and {Marchese}, Elijah J. and {Stanford}, S. Adam and {Caselden}, Dan and {Cushing}, Michael C. and {Cutri}, Roc M. and {Faherty}, Jacqueline K. and {Gelino}, Christopher R. and {Gonzalez}, Anthony H. and {Mainzer}, Amanda and {Mobasher}, Bahram and {Schlegel}, David J. and {Stern}, Daniel and {Teplitz}, Harry I. and {Wright}, Edward L.},
        title = "{The CatWISE Preliminary Catalog: Motions from WISE and NEOWISE Data}",
      journal = {\apjs},
         year = 2020,
        month = apr,
       volume = {247},
       number = {2},
          eid = {69},
        pages = {69},
          doi = {10.3847/1538-4365/ab7f2a},
archivePrefix = {arXiv},
       eprint = {1908.08902},
 primaryClass = {astro-ph.IM},
       adsurl = {https://ui.adsabs.harvard.edu/abs/2020ApJS..247...69E}
}

@ARTICLE{Rasia1509.04247,
       author = {{Rasia}, E. and {Borgani}, S. and {Murante}, G. and {Planelles}, S. and {Beck}, A.~M. and {Biffi}, V. and {Ragone-Figueroa}, C. and {Granato}, G.~L. and {Steinborn}, L.~K. and {Dolag}, K.},
        title = "{Cool Core Clusters from Cosmological Simulations}",
      journal = {\apjl},
         year = 2015,
        month = nov,
       volume = {813},
       number = {1},
          eid = {L17},
        pages = {L17},
          doi = {10.1088/2041-8205/813/1/L17},
archivePrefix = {arXiv},
       eprint = {1509.04247},
 primaryClass = {astro-ph.CO},
       adsurl = {https://ui.adsabs.harvard.edu/abs/2015ApJ...813L..17R}
}

@ARTICLE{De-Grandi0012232,
       author = {{De Grandi}, Sabrina and {Molendi}, Silvano},
        title = "{Metallicity Gradients in X-Ray Clusters of Galaxies}",
      journal = {\apj},
         year = 2001,
        month = apr,
       volume = {551},
       number = {1},
        pages = {153-159},
          doi = {10.1086/320098},
archivePrefix = {arXiv},
       eprint = {astro-ph/0012232},
 primaryClass = {astro-ph},
       adsurl = {https://ui.adsabs.harvard.edu/abs/2001ApJ...551..153D}
}

@ARTICLE{Smith1406.2037,
       author = {{Smith}, Randall K. and {Foster}, Adam R. and {Edgar}, Richard J. and {Brickhouse}, Nancy S.},
        title = "{Resolving the Origin of the Diffuse Soft X-Ray Background}",
      journal = {\apj},
         year = 2014,
        month = may,
       volume = {787},
       number = {1},
          eid = {77},
        pages = {77},
          doi = {10.1088/0004-637X/787/1/77},
archivePrefix = {arXiv},
       eprint = {1406.2037},
 primaryClass = {astro-ph.HE},
       adsurl = {https://ui.adsabs.harvard.edu/abs/2014ApJ...787...77S}
}

@ARTICLE{Snowden1995ApJ...454..643S,
       author = {{Snowden}, S.~L. and {Freyberg}, M.~J. and {Plucinsky}, P.~P. and {Schmitt}, J.~H.~M.~M. and {Truemper}, J. and {Voges}, W. and {Edgar}, R.~J. and {McCammon}, D. and {Sanders}, W.~T.},
        title = "{First Maps of the Soft X-Ray Diffuse Background from the ROSAT XRT/PSPC All-Sky Survey}",
      journal = {\apj},
         year = 1995,
        month = dec,
       volume = {454},
        pages = {643},
          doi = {10.1086/176517},
       adsurl = {https://ui.adsabs.harvard.edu/abs/1995ApJ...454..643S}
}

@ARTICLE{Snowden1997ApJ...485..125S,
       author = {{Snowden}, S.~L. and {Egger}, R. and {Freyberg}, M.~J. and {McCammon}, D. and {Plucinsky}, P.~P. and {Sanders}, W.~T. and {Schmitt}, J.~H.~M.~M. and {Tr{\"u}mper}, J. and {Voges}, W.},
        title = "{ROSAT Survey Diffuse X-Ray Background Maps. II.}",
      journal = {\apj},
         year = 1997,
        month = aug,
       volume = {485},
       number = {1},
        pages = {125-135},
          doi = {10.1086/304399},
       adsurl = {https://ui.adsabs.harvard.edu/abs/1997ApJ...485..125S}
}

@article{Suzuki2108.11234,
       author = {{Suzuki}, H. and {Plucinsky}, P.~P. and {Gaetz}, T.~J. and {Bamba}, A.},
        title = "{Spatial and temporal variations of the Chandra ACIS particle-induced background and development of a spectral-model generation tool}",
      journal = {\aap},
         year = 2021,
        month = nov,
       volume = {655},
          eid = {A116},
        pages = {A116},
          doi = {10.1051/0004-6361/202141458},
archivePrefix = {arXiv},
       eprint = {2108.11234},
 primaryClass = {astro-ph.HE},
       adsurl = {https://ui.adsabs.harvard.edu/abs/2021A&A...655A.116S}
}

@article{Dey1804.08657,
       author = {{Dey}, Arjun and {Schlegel}, David J. and {Lang}, Dustin and {Blum}, Robert and {Burleigh}, Kaylan and {Fan}, Xiaohui and {Findlay}, Joseph R. and {Finkbeiner}, Doug and {Herrera}, David and {Juneau}, St{\'e}phanie and {Landriau}, Martin and {Levi}, Michael and {McGreer}, Ian and {Meisner}, Aaron and {Myers}, Adam D. and {Moustakas}, John and {Nugent}, Peter and {Patej}, Anna and {Schlafly}, Edward F. and {Walker}, Alistair R. and {Valdes}, Francisco and {Weaver}, Benjamin A. and {Y{\`e}che}, Christophe and {Zou}, Hu and {Zhou}, Xu and {Abareshi}, Behzad and {Abbott}, T.~M.~C. and {Abolfathi}, Bela and {Aguilera}, C. and {Alam}, Shadab and {Allen}, Lori and {Alvarez}, A. and {Annis}, James and {Ansarinejad}, Behzad and {Aubert}, Marie and {Beechert}, Jacqueline and {Bell}, Eric F. and {BenZvi}, Segev Y. and {Beutler}, Florian and {Bielby}, Richard M. and {Bolton}, Adam S. and {Brice{\~n}o}, C{\'e}sar and {Buckley-Geer}, Elizabeth J. and {Butler}, Karen and {Calamida}, Annalisa and {Carlberg}, Raymond G. and {Carter}, Paul and {Casas}, Ricard and {Castander}, Francisco J. and {Choi}, Yumi and {Comparat}, Johan and {Cukanovaite}, Elena and {Delubac}, Timoth{\'e}e and {DeVries}, Kaitlin and {Dey}, Sharmila and {Dhungana}, Govinda and {Dickinson}, Mark and {Ding}, Zhejie and {Donaldson}, John B. and {Duan}, Yutong and {Duckworth}, Christopher J. and {Eftekharzadeh}, Sarah and {Eisenstein}, Daniel J. and {Etourneau}, Thomas and {Fagrelius}, Parker A. and {Farihi}, Jay and {Fitzpatrick}, Mike and {Font-Ribera}, Andreu and {Fulmer}, Leah and {G{\"a}nsicke}, Boris T. and {Gaztanaga}, Enrique and {George}, Koshy and {Gerdes}, David W. and {Gontcho}, Satya Gontcho A. and {Gorgoni}, Claudio and {Green}, Gregory and {Guy}, Julien and {Harmer}, Diane and {Hernandez}, M. and {Honscheid}, Klaus and {Huang}, Lijuan Wendy and {James}, David J. and {Jannuzi}, Buell T. and {Jiang}, Linhua and {Joyce}, Richard and {Karcher}, Armin and {Karkar}, Sonia and {Kehoe}, Robert and {Kneib}, Jean-Paul and {Kueter-Young}, Andrea and {Lan}, Ting-Wen and {Lauer}, Tod R. and {Le Guillou}, Laurent and {Le Van Suu}, Auguste and {Lee}, Jae Hyeon and {Lesser}, Michael and {Perreault Levasseur}, Laurence and {Li}, Ting S. and {Mann}, Justin L. and {Marshall}, Robert and {Mart{\'\i}nez-V{\'a}zquez}, C.~E. and {Martini}, Paul and {du Mas des Bourboux}, H{\'e}lion and {McManus}, Sean and {Meier}, Tobias Gabriel and {M{\'e}nard}, Brice and {Metcalfe}, Nigel and {Mu{\~n}oz-Guti{\'e}rrez}, Andrea and {Najita}, Joan and {Napier}, Kevin and {Narayan}, Gautham and {Newman}, Jeffrey A. and {Nie}, Jundan and {Nord}, Brian and {Norman}, Dara J. and {Olsen}, Knut A.~G. and {Paat}, Anthony and {Palanque-Delabrouille}, Nathalie and {Peng}, Xiyan and {Poppett}, Claire L. and {Poremba}, Megan R. and {Prakash}, Abhishek and {Rabinowitz}, David and {Raichoor}, Anand and {Rezaie}, Mehdi and {Robertson}, A.~N. and {Roe}, Natalie A. and {Ross}, Ashley J. and {Ross}, Nicholas P. and {Rudnick}, Gregory and {Safonova}, Sasha and {Saha}, Abhijit and {S{\'a}nchez}, F. Javier and {Savary}, Elodie and {Schweiker}, Heidi and {Scott}, Adam and {Seo}, Hee-Jong and {Shan}, Huanyuan and {Silva}, David R. and {Slepian}, Zachary and {Soto}, Christian and {Sprayberry}, David and {Staten}, Ryan and {Stillman}, Coley M. and {Stupak}, Robert J. and {Summers}, David L. and {Sien Tie}, Suk and {Tirado}, H. and {Vargas-Maga{\~n}a}, Mariana and {Vivas}, A. Katherina and {Wechsler}, Risa H. and {Williams}, Doug and {Yang}, Jinyi and {Yang}, Qian and {Yapici}, Tolga and {Zaritsky}, Dennis and {Zenteno}, A. and {Zhang}, Kai and {Zhang}, Tianmeng and {Zhou}, Rongpu and {Zhou}, Zhimin},
        title = "{Overview of the DESI Legacy Imaging Surveys}",
      journal = {\aj},
         year = 2019,
        month = may,
       volume = {157},
       number = {5},
          eid = {168},
        pages = {168},
          doi = {10.3847/1538-3881/ab089d},
archivePrefix = {arXiv},
       eprint = {1804.08657},
 primaryClass = {astro-ph.IM},
       adsurl = {https://ui.adsabs.harvard.edu/abs/2019AJ....157..168D}
}

@ARTICLE{Darragh-Ford2302.10931,
       author = {{Darragh-Ford}, Elise and {Mantz}, Adam B. and {Rasia}, Elena and {Allen}, Steven W. and {Morris}, R. Glenn and {Foster}, Jack and {Schmidt}, Robert W. and {Wenrich}, Guillermo},
        title = "{The Concentration-Mass relation of massive, dynamically relaxed galaxy clusters: agreement between observations and {\ensuremath{\Lambda}}CDM simulations}",
      journal = {\mnras},
         year = 2023,
        month = may,
       volume = {521},
       number = {1},
        pages = {790-799},
          doi = {10.1093/mnras/stad585},
archivePrefix = {arXiv},
       eprint = {2302.10931},
 primaryClass = {astro-ph.CO},
       adsurl = {https://ui.adsabs.harvard.edu/abs/2023MNRAS.521..790D}
}

@ARTICLE{Calzadilla2303.10185,
       author = {{Calzadilla}, Michael S. and {Bleem}, Lindsey E. and {McDonald}, Michael and {Gladders}, Michael D. and {Mantz}, Adam B. and {Allen}, Steven W. and {Bayliss}, Matthew B. and {Eilers}, Anna-Christina and {Floyd}, Benjamin and {Hlavacek-Larrondo}, Julie and {Khullar}, Gourav and {Kim}, Keunho J. and {Mahler}, Guillaume and {Sharon}, Keren and {Somboonpanyakul}, Taweewat and {Stalder}, Brian and {Stark}, Antony A. and {SPT Collaboration}},
        title = "{SPT-CL J2215-3537: A Massive Starburst at the Center of the Most Distant Relaxed Galaxy Cluster}",
      journal = {\apj},
         year = 2023,
        month = apr,
       volume = {947},
       number = {2},
          eid = {44},
        pages = {44},
          doi = {10.3847/1538-4357/acc6c2},
archivePrefix = {arXiv},
       eprint = {2303.10185},
 primaryClass = {astro-ph.GA},
       adsurl = {https://ui.adsabs.harvard.edu/abs/2023ApJ...947...44C}
}

@ARTICLE{Somboonpanyakul2201.08398,
       author = {{Somboonpanyakul}, T. and {McDonald}, M. and {Noble}, A. and {Aguena}, M. and {Allam}, S. and {Amon}, A. and {Andrade-Oliveira}, F. and {Bacon}, D. and {Bayliss}, M.~B. and {Bertin}, E. and {Bhargava}, S. and {Brooks}, D. and {Buckley-Geer}, E. and {Burke}, D.~L. and {Calzadilla}, M. and {Canning}, R. and {Carnero Rosell}, A. and {Carrasco Kind}, M. and {Carretero}, J. and {Costanzi}, M. and {da Costa}, L.~N. and {Pereira}, M.~E.~S. and {De Vicente}, J. and {Doel}, P. and {Eisenhardt}, P. and {Everett}, S. and {Evrard}, A.~E. and {Ferrero}, I. and {Flaugher}, B. and {Floyd}, B. and {Garc{\'\i}a-Bellido}, J. and {Gaztanaga}, E. and {Gerdes}, D.~W. and {Gonzalez}, A. and {Gruen}, D. and {Gruendl}, R.~A. and {Gschwend}, J. and {Gupta}, N. and {Gutierrez}, G. and {Hinton}, S.~R. and {Hollowood}, D.~L. and {Honscheid}, K. and {Hoyle}, B. and {James}, D.~J. and {Jeltema}, T. and {Khullar}, G. and {Kim}, K.~J. and {Klein}, M. and {Kuehn}, K. and {Lima}, M. and {Maia}, M.~A.~G. and {Marshall}, J.~L. and {Martini}, P. and {Melchior}, P. and {Menanteau}, F. and {Miquel}, R. and {Mohr}, J.~J. and {Morgan}, R. and {Ogando}, R.~L.~C. and {Palmese}, A. and {Paz-Chinch{\'o}n}, F. and {Pieres}, A. and {Plazas Malag{\'o}n}, A.~A. and {Reil}, K. and {Romer}, A.~K. and {Ruppin}, F. and {Sanchez}, E. and {Saro}, A. and {Scarpine}, V. and {Schubnell}, M. and {Serrano}, S. and {Sevilla-Noarbe}, I. and {Singh}, P. and {Smith}, M. and {Soares-Santos}, M. and {Strazzullo}, V. and {Suchyta}, E. and {Swanson}, M.~E.~C. and {Tarle}, G. and {To}, C. and {Tucker}, D.~L. and {Wilkinson}, R.~D.},
        title = "{The Evolution of AGN Activity in Brightest Cluster Galaxies}",
      journal = {\aj},
         year = 2022,
        month = apr,
       volume = {163},
       number = {4},
          eid = {146},
        pages = {146},
          doi = {10.3847/1538-3881/ac5030},
archivePrefix = {arXiv},
       eprint = {2201.08398},
 primaryClass = {astro-ph.GA},
       adsurl = {https://ui.adsabs.harvard.edu/abs/2022AJ....163..146S}
}

@article{Ruppin2012.14669,
       author = {{Ruppin}, F. and {McDonald}, M. and {Bleem}, L.~E. and {Allen}, S.~W. and {Benson}, B.~A. and {Calzadilla}, M. and {Khullar}, G. and {Floyd}, B.},
        title = "{Stability of Cool Cores during Galaxy Cluster Growth: A Joint Chandra/SPT Analysis of 67 Galaxy Clusters along a Common Evolutionary Track Spanning 9 Gyr}",
      journal = {\apj},
         year = 2021,
        month = sep,
       volume = {918},
       number = {2},
          eid = {43},
        pages = {43},
          doi = {10.3847/1538-4357/ac0bba},
archivePrefix = {arXiv},
       eprint = {2012.14669},
 primaryClass = {astro-ph.CO},
       adsurl = {https://ui.adsabs.harvard.edu/abs/2021ApJ...918...43R}
}

@article{Mantz2111.09343,
	adsurl = {https://ui.adsabs.harvard.edu/abs/2022MNRAS.510..131M},
	archiveprefix = {arXiv},
	author = {{Mantz}, Adam B. and {Morris}, R. Glenn and {Allen}, Steven W. and {Canning}, Rebecca E.~A. and {Baumont}, Lucie and {Benson}, Bradford and {Bleem}, Lindsey E. and {Ehlert}, Steven R. and {Floyd}, Benjamin and {Herbonnet}, Ricardo and {Kelly}, Patrick L. and {Liang}, Shuang and {von der Linden}, Anja and {McDonald}, Michael and {Rapetti}, David A. and {Schmidt}, Robert W. and {Werner}, Norbert and {Wright}, Adam},
	doi = {10.1093/mnras/stab3390},
	eprint = {2111.09343},
	journal = {\mnras},
	month = feb,
	number = {1},
	pages = {131-145},
	primaryclass = {astro-ph.CO},
	title = {{Cosmological constraints from gas mass fractions of massive, relaxed galaxy clusters}},
	volume = {510},
	year = 2022}

@article{Flores2108.12051,
	adsurl = {https://ui.adsabs.harvard.edu/abs/2021MNRAS.507.5195F},
	archiveprefix = {arXiv},
	author = {{Flores}, Anthony M. and {Mantz}, Adam B. and {Allen}, Steven W. and {Morris}, R. Glenn and {Canning}, Rebecca E.~A. and {Bleem}, Lindsey E. and {Calzadilla}, Michael S. and {Floyd}, Benjamin T. and {McDonald}, Michael and {Ruppin}, Florian},
	doi = {10.1093/mnras/stab2430},
	eprint = {2108.12051},
	journal = {\mnras},
	month = nov,
	number = {4},
	pages = {5195-5204},
	primaryclass = {astro-ph.CO},
	title = {{The history of metal enrichment traced by X-ray observations of high-redshift galaxy clusters}},
	volume = {507},
	year = 2021}

@article{Wan2101.09389,
	adsurl = {https://ui.adsabs.harvard.edu/abs/2021MNRAS.504.1062W},
	archiveprefix = {arXiv},
	author = {{Wan}, Jenny T. and {Mantz}, Adam B. and {Sayers}, Jack and {Allen}, Steven W. and {Morris}, R. Glenn and {Golwala}, Sunil R.},
	doi = {10.1093/mnras/stab948},
	eprint = {2101.09389},
	journal = {\mnras},
	month = jun,
	number = {1},
	pages = {1062-1076},
	primaryclass = {astro-ph.CO},
	title = {{Measuring H$_{0}$ using X-ray and SZ effect observations of dynamically relaxed galaxy clusters}},
	volume = {504},
	year = 2021}

@article{Huang1907.09621,
	adsurl = {https://ui.adsabs.harvard.edu/abs/2020AJ....159..110H},
	archiveprefix = {arXiv},
	author = {{Huang}, N. and {Bleem}, L.~E. and {Stalder}, B. and {Ade}, P.~A.~R. and {Allen}, S.~W. and {Anderson}, A.~J. and {Austermann}, J.~E. and {Avva}, J.~S. and {Beall}, J.~A. and {Bender}, A.~N. and {Benson}, B.~A. and {Bianchini}, F. and {Bocquet}, S. and {Brodwin}, M. and {Carlstrom}, J.~E. and {Chang}, C.~L. and {Chiang}, H.~C. and {Citron}, R. and {Moran}, C. Corbett and {Crawford}, T.~M. and {Crites}, A.~T. and {Haan}, T. de and {Dobbs}, M.~A. and {Everett}, W. and {Floyd}, B. and {Gallicchio}, J. and {George}, E.~M. and {Gilbert}, A. and {Gladders}, M.~D. and {Guns}, S. and {Gupta}, N. and {Halverson}, N.~W. and {Harrington}, N. and {Henning}, J.~W. and {Hilton}, G.~C. and {Holder}, G.~P. and {Holzapfel}, W.~L. and {Hrubes}, J.~D. and {Hubmayr}, J. and {Irwin}, K.~D. and {Khullar}, G. and {Knox}, L. and {Lee}, A.~T. and {Li}, D. and {Lowitz}, A. and {McDonald}, M. and {McMahon}, J.~J. and {Meyer}, S.~S. and {Mocanu}, L.~M. and {Montgomery}, J. and {Nadolski}, A. and {Natoli}, T. and {Nibarger}, J.~P. and {Noble}, G. and {Novosad}, V. and {Padin}, S. and {Patil}, S. and {Pryke}, C. and {Reichardt}, C.~L. and {Ruhl}, J.~E. and {Saliwanchik}, B.~R. and {Saro}, A. and {Sayre}, J.~T. and {Schaffer}, K.~K. and {Sharon}, K. and {Sievers}, C. and {Smecher}, G. and {Stark}, A.~A. and {Story}, K.~T. and {Tucker}, C. and {Vanderlinde}, K. and {Veach}, T. and {Vieira}, J.~D. and {Wang}, G. and {Whitehorn}, N. and {Wu}, W.~L.~K. and {Yefremenko}, V.},
	doi = {10.3847/1538-3881/ab6a96},
	eid = {110},
	eprint = {1907.09621},
	journal = {\aj},
	month = mar,
	number = {3},
	pages = {110},
	primaryclass = {astro-ph.CO},
	title = {{Galaxy Clusters Selected via the Sunyaev-Zel'dovich Effect in the SPTpol 100-square-degree Survey}},
	volume = {159},
	year = 2020}

@article{Hilton2009.11043,
	adsurl = {https://ui.adsabs.harvard.edu/abs/2021ApJS..253....3H},
	archiveprefix = {arXiv},
	author = {{Hilton}, M. and {Sif{\'o}n}, C. and {Naess}, S. and {Madhavacheril}, M. and {Oguri}, M. and {Rozo}, E. and {Rykoff}, E. and {Abbott}, T.~M.~C. and {Adhikari}, S. and {Aguena}, M. and {Aiola}, S. and {Allam}, S. and {Amodeo}, S. and {Amon}, A. and {Annis}, J. and {Ansarinejad}, B. and {Aros-Bunster}, C. and {Austermann}, J.~E. and {Avila}, S. and {Bacon}, D. and {Battaglia}, N. and {Beall}, J.~A. and {Becker}, D.~T. and {Bernstein}, G.~M. and {Bertin}, E. and {Bhandarkar}, T. and {Bhargava}, S. and {Bond}, J.~R. and {Brooks}, D. and {Burke}, D.~L. and {Calabrese}, E. and {Carrasco Kind}, M. and {Carretero}, J. and {Choi}, S.~K. and {Choi}, A. and {Conselice}, C. and {da Costa}, L.~N. and {Costanzi}, M. and {Crichton}, D. and {Crowley}, K.~T. and {D{\"u}nner}, R. and {Denison}, E.~V. and {Devlin}, M.~J. and {Dicker}, S.~R. and {Diehl}, H.~T. and {Dietrich}, J.~P. and {Doel}, P. and {Duff}, S.~M. and {Duivenvoorden}, A.~J. and {Dunkley}, J. and {Everett}, S. and {Ferraro}, S. and {Ferrero}, I. and {Fert{\'e}}, A. and {Flaugher}, B. and {Frieman}, J. and {Gallardo}, P.~A. and {Garc{\'\i}a-Bellido}, J. and {Gaztanaga}, E. and {Gerdes}, D.~W. and {Giles}, P. and {Golec}, J.~E. and {Gralla}, M.~B. and {Grandis}, S. and {Gruen}, D. and {Gruendl}, R.~A. and {Gschwend}, J. and {Gutierrez}, G. and {Han}, D. and {Hartley}, W.~G. and {Hasselfield}, M. and {Hill}, J.~C. and {Hilton}, G.~C. and {Hincks}, A.~D. and {Hinton}, S.~R. and {Ho}, S. -P.~P. and {Honscheid}, K. and {Hoyle}, B. and {Hubmayr}, J. and {Huffenberger}, K.~M. and {Hughes}, J.~P. and {Jaelani}, A.~T. and {Jain}, B. and {James}, D.~J. and {Jeltema}, T. and {Kent}, S. and {Knowles}, K. and {Koopman}, B.~J. and {Kuehn}, K. and {Lahav}, O. and {Lima}, M. and {Lin}, Y. -T. and {Lokken}, M. and {Loubser}, S.~I. and {MacCrann}, N. and {Maia}, M.~A.~G. and {Marriage}, T.~A. and {Martin}, J. and {McMahon}, J. and {Melchior}, P. and {Menanteau}, F. and {Miquel}, R. and {Miyatake}, H. and {Moodley}, K. and {Morgan}, R. and {Mroczkowski}, T. and {Nati}, F. and {Newburgh}, L.~B. and {Niemack}, M.~D. and {Nishizawa}, A.~J. and {Ogando}, R.~L.~C. and {Orlowski-Scherer}, J. and {Page}, L.~A. and {Palmese}, A. and {Partridge}, B. and {Paz-Chinch{\'o}n}, F. and {Phakathi}, P. and {Plazas}, A.~A. and {Robertson}, N.~C. and {Romer}, A.~K. and {Carnero Rosell}, A. and {Salatino}, M. and {Sanchez}, E. and {Schaan}, E. and {Schillaci}, A. and {Sehgal}, N. and {Serrano}, S. and {Shin}, T. and {Simon}, S.~M. and {Smith}, M. and {Soares-Santos}, M. and {Spergel}, D.~N. and {Staggs}, S.~T. and {Storer}, E.~R. and {Suchyta}, E. and {Swanson}, M.~E.~C. and {Tarle}, G. and {Thomas}, D. and {To}, C. and {Trac}, H. and {Ullom}, J.~N. and {Vale}, L.~R. and {Van Lanen}, J. and {Vavagiakis}, E.~M. and {De Vicente}, J. and {Wilkinson}, R.~D. and {Wollack}, E.~J. and {Xu}, Z. and {Zhang}, Y.},
	doi = {10.3847/1538-4365/abd023},
	eid = {3},
	eprint = {2009.11043},
	journal = {\apjs},
	month = mar,
	number = {1},
	pages = {3},
	primaryclass = {astro-ph.CO},
	title = {{The Atacama Cosmology Telescope: A Catalog of >4000 Sunyaev{\textendash}Zel{\textquoteright}dovich Galaxy Clusters}},
	volume = {253},
	year = 2021}

@ARTICLE{Smith0106478,
       author = {{Smith}, Randall K. and {Brickhouse}, Nancy S. and {Liedahl}, Duane A. and {Raymond}, John C.},
        title = "{Collisional Plasma Models with APEC/APED: Emission-Line Diagnostics of Hydrogen-like and Helium-like Ions}",
      journal = {\apjl},
         year = 2001,
        month = aug,
       volume = {556},
       number = {2},
        pages = {L91-L95},
          doi = {10.1086/322992},
archivePrefix = {arXiv},
       eprint = {astro-ph/0106478},
 primaryClass = {astro-ph},
       adsurl = {https://ui.adsabs.harvard.edu/abs/2001ApJ...556L..91S}
}

@article{Miyaji1503.00056,
	adsurl = {https://ui.adsabs.harvard.edu/abs/2015ApJ...804..104M},
	archiveprefix = {arXiv},
	author = {{Miyaji}, T. and {Hasinger}, G. and {Salvato}, M. and {Brusa}, M. and {Cappelluti}, N. and {Civano}, F. and {Puccetti}, S. and {Elvis}, M. and {Brunner}, H. and {Fotopoulou}, S. and {Ueda}, Y. and {Griffiths}, R.~E. and {Koekemoer}, A.~M. and {Akiyama}, M. and {Comastri}, A. and {Gilli}, R. and {Lanzuisi}, G. and {Merloni}, A. and {Vignali}, C.},
	doi = {10.1088/0004-637X/804/2/104},
	eid = {104},
	eprint = {1503.00056},
	journal = {\apj},
	month = may,
	number = {2},
	pages = {104},
	primaryclass = {astro-ph.GA},
	title = {{Detailed Shape and Evolutionary Behavior of the X-Ray Luminosity Function of Active Galactic Nuclei}},
	volume = {804},
	year = 2015}

@article{Bleem1910.04121,
	adsurl = {https://ui.adsabs.harvard.edu/abs/2020ApJS..247...25B},
	archiveprefix = {arXiv},
	author = {{Bleem}, L.~E. and {Bocquet}, S. and {Stalder}, B. and {Gladders}, M.~D. and {Ade}, P.~A.~R. and {Allen}, S.~W. and {Anderson}, A.~J. and {Annis}, J. and {Ashby}, M.~L.~N. and {Austermann}, J.~E. and {Avila}, S. and {Avva}, J.~S. and {Bayliss}, M. and {Beall}, J.~A. and {Bechtol}, K. and {Bender}, A.~N. and {Benson}, B.~A. and {Bertin}, E. and {Bianchini}, F. and {Blake}, C. and {Brodwin}, M. and {Brooks}, D. and {Buckley-Geer}, E. and {Burke}, D.~L. and {Carlstrom}, J.~E. and {Rosell}, A. Carnero and {Carrasco Kind}, M. and {Carretero}, J. and {Chang}, C.~L. and {Chiang}, H.~C. and {Citron}, R. and {Moran}, C. Corbett and {Costanzi}, M. and {Crawford}, T.~M. and {Crites}, A.~T. and {da Costa}, L.~N. and {de Haan}, T. and {De Vicente}, J. and {Desai}, S. and {Diehl}, H.~T. and {Dietrich}, J.~P. and {Dobbs}, M.~A. and {Eifler}, T.~F. and {Everett}, W. and {Flaugher}, B. and {Floyd}, B. and {Frieman}, J. and {Gallicchio}, J. and {Garc{\'\i}a-Bellido}, J. and {George}, E.~M. and {Gerdes}, D.~W. and {Gilbert}, A. and {Gruen}, D. and {Gruendl}, R.~A. and {Gschwend}, J. and {Gupta}, N. and {Gutierrez}, G. and {Halverson}, N.~W. and {Harrington}, N. and {Henning}, J.~W. and {Heymans}, C. and {Holder}, G.~P. and {Hollowood}, D.~L. and {Holzapfel}, W.~L. and {Honscheid}, K. and {Hrubes}, J.~D. and {Huang}, N. and {Hubmayr}, J. and {Irwin}, K.~D. and {James}, D.~J. and {Jeltema}, T. and {Joudaki}, S. and {Khullar}, G. and {Klein}, M. and {Knox}, L. and {Kuropatkin}, N. and {Lee}, A.~T. and {Li}, D. and {Lidman}, C. and {Lowitz}, A. and {MacCrann}, N. and {Mahler}, G. and {Maia}, M.~A.~G. and {Marshall}, J.~L. and {McDonald}, M. and {McMahon}, J.~J. and {Melchior}, P. and {Menanteau}, F. and {Meyer}, S.~S. and {Miquel}, R. and {Mocanu}, L.~M. and {Mohr}, J.~J. and {Montgomery}, J. and {Nadolski}, A. and {Natoli}, T. and {Nibarger}, J.~P. and {Noble}, G. and {Novosad}, V. and {Padin}, S. and {Palmese}, A. and {Parkinson}, D. and {Patil}, S. and {Paz-Chinch{\'o}n}, F. and {Plazas}, A.~A. and {Pryke}, C. and {Ramachandra}, N.~S. and {Reichardt}, C.~L. and {Remolina Gonz{\'a}lez}, J.~D. and {Romer}, A.~K. and {Roodman}, A. and {Ruhl}, J.~E. and {Rykoff}, E.~S. and {Saliwanchik}, B.~R. and {Sanchez}, E. and {Saro}, A. and {Sayre}, J.~T. and {Schaffer}, K.~K. and {Schrabback}, T. and {Serrano}, S. and {Sharon}, K. and {Sievers}, C. and {Smecher}, G. and {Smith}, M. and {Soares-Santos}, M. and {Stark}, A.~A. and {Story}, K.~T. and {Suchyta}, E. and {Tarle}, G. and {Tucker}, C. and {Vanderlinde}, K. and {Veach}, T. and {Vieira}, J.~D. and {Wang}, G. and {Weller}, J. and {Whitehorn}, N. and {Wu}, W.~L.~K. and {Yefremenko}, V. and {Zhang}, Y.},
	doi = {10.3847/1538-4365/ab6993},
	eid = {25},
	eprint = {1910.04121},
	journal = {\apjs},
	month = mar,
	number = {1},
	pages = {25},
	primaryclass = {astro-ph.CO},
	title = {{The SPTpol Extended Cluster Survey}},
	volume = {247},
	year = 2020}
\bibliographystyle{aasjournal}



\end{document}